\documentclass[pdflatex,sn-mathphys-num]{sn-jnl}% Math and Physical Sciences Numbered Reference Style 
\usepackage{graphicx}%
\usepackage{multirow}%
\usepackage{amsmath,amssymb,amsfonts}%
\usepackage{amsthm}%
\usepackage{mathrsfs}%
\usepackage[title]{appendix}%
\usepackage[table]{xcolor}%
\usepackage{textcomp}%
\usepackage{manyfoot}%
\usepackage{booktabs}%
\usepackage{algorithm}%
\usepackage{algorithmicx}%
\usepackage{algpseudocode}%
\usepackage{listings}%
\usepackage{array}
\usepackage[displaymath, mathlines]{lineno}

\theoremstyle{thmstyleone}%
\theoremstyle{thmstyletwo}%

\theoremstyle{thmstylethree}%

\begin{document}

\title[A theoretical analysis of evolvability]{First explore, then settle: a theoretical analysis of evolvability as a driver of adaptation}


\author*[1,2]{\fnm{Juan} \sur{Jim\'{e}nez-S\'{a}nchez}}\email{Juan.JSanchez@uclm.es}

\author[2]{\fnm{Carmen} \sur{Ortega-Sabater}}%\email{carmen.ortega@uclm.es}

\author[3]{\fnm{Philip K.} \sur{Maini}}%\email{victor.perezgarcia@uclm.es}

\author[2]{\fnm{V\'{i}ctor M.} \sur{P\'{e}rez-Garc\'{i}a}}%\email{philip.maini@maths.ox.ac.uk}

\author[1]{\fnm{Tommaso} \sur{Lorenzi}}%\email{tommaso.lorenzi@polito.it}

\affil*[1]{{\small \orgdiv{Dipartimento di Scienze Matematiche (DISMA)}, \orgname{Politecnico di Torino}, \orgaddress{\street{Corso Duca degli Abruzzi 24}, \city{Turin}, \postcode{10129}, \country{Italy}}}}

\affil[2]{{\small\orgdiv{Mathematical Oncology Laboratory (MOLAB)}, \orgname{Universidad de Castilla-La Mancha (UCLM)}, \orgaddress{\street{Avda. Camilo Jos\'{e} Cela}, \city{Ciudad Real}, \postcode{13071}, \country{Spain}}}}

\affil[3]{{\small\orgdiv{Wolfson Centre for Mathematical Biology}, \orgname{Mathematical Institute, University of Oxford}, \orgaddress{\street{Woodstock Road}, \city{Oxford}, \postcode{OX2 6GG},  \country{United Kingdom}}}}


\abstract{Evolvability is defined as the ability of a population to generate heritable variation to facilitate its adaptation to new environments or selection pressures. In this article, we consider evolvability as a phenotypic trait subject to evolution and discuss its implications in the adaptation of populations of asexual individuals. We explore the evolutionary dynamics of an actively proliferating population of individuals, subject to changes in their proliferative potential and their evolvability, through mathematical simulations of a stochastic individual-based model and its deterministic continuum counterpart. We find robust adaptive trajectories that rely on individuals with high evolvability rapidly exploring the phenotypic landscape and reaching the proliferative potential with the highest fitness. The strength of selection on the proliferative potential, and the cost associated with evolvability, can alter these trajectories such that, if both are sufficiently constraining, highly evolvable populations can become extinct in our individual-based model simulations. We explore the impact of this interaction at various scales, discussing its effects in undisturbed environments and also in disrupted contexts, such as cancer.}




\maketitle

%\linenumbers

\section{Introduction}
\label{sec1}

In nature, variation in phenotypic traits and selection are the general conditions for evolution to occur \cite{darwin1859,houle1992,lynch1998,hoffmann1999}. Selection acts on phenotypic variants to promote the prevalence of the fittest. The greater the diversity within a population, the higher its chances of withstanding the pressures imposed by selective agents \cite{barrett2008}. Phenotypic variability is generated by a plethora of biological mechanisms, most of which operate at the cellular level, but whose effects are manifested at the macroscopic level \cite{sigal2006, whiting2024}. Arguably one of the mechanisms that operates at the most fundamental level is mutation. The stochastic nature of mutations, as well as their high occurrence \cite{kunkel2000}, results in the generation of both detrimental and beneficial variants. According to recent studies on the distribution of fitness effects, although detrimental variants are more frequent \cite{eyre2007}, beneficial and neutral mutations must be present for adaptation to occur \cite{eyre2006}.
Even though observed cell-to-cell variability has traditionally been ascribed to genetic \cite{galhardo2007} and epigenetic variation \cite{true2004}, the stochastic nature of cellular processes \cite{blake2003,elowitz2002, allan2009} brings into play a non-genetic heterogeneity \cite{capp2021, sigal2006, mcadams1999,raser2005,sanchez2013}. Non-genetic heterogeneity is also responsible for the emergence of phenotypic variation \cite{mtguinn2020} and manifests itself as variations in cell size, function, lifespan, protein level and more.
This ability to generate variability on which selection can act upon is known as evolvability \cite{c4_evol1}.

More formally, evolvability can be defined as a population's capacity to provide adaptive and heritable phenotypic variation amongst its individuals, to let them evolve and overcome selective pressures. The above definition could be considered a consensus of overlapping ideas amongst different evolutionary biologists, as each brings a different nuance depending on their concept of evolvability and the scale at which they study it \cite{c4_evol1,c4_evol2,c4_evol3,c4_evol4}. It could be argued that greater evolvability implies greater selective advantage, as the adaptive potential increases the likelihood of surviving selective pressures. However, different ecological scenarios may foster populations with lower evolvability \cite{bukkuri2023}, especially under environmental stasis. Higher evolvability is not free of charge, since it may also boost the generation of detrimental phenotypes \cite{cohen1966,seger1987,slatkin1974}.

Evolvability itself may be evolvable and subject to change over time \cite{c4_evol3}, suggesting that it should also be considered as a phenotypic trait \cite{riederer2022}. 
A balance between robustness and evolvability is required so that new phenotypic states can be explored without detrimentally affecting essential features \cite{c4_evol2}. This relationship between robustness and evolvability has, as an example at the molecular level, been already characterised in the context of transcription factor binding sites in mice and yeast \cite{payne2014}, and evolvability-enhancing mutations were recently described in the context of RNA and proteins \cite{wagner2023}.

Alterations that increase mutation rates \cite{colegrave2008} and phenotypic variation \cite{bodi2017} may promote evolvability, as has been described in bacteria and yeast \cite{bodi2017, wagner2023}. Hence, the genetic pathways that are involved in genomic repair and instability, recombination, horizontal gene transfer and regulation of gene expression are likely to influence the level of evolvability. Moreover, some theoretical studies have addressed the evolution of evolvability in gene regulatory networks \cite{crombach2008,draghi2009}.

On a broader scale, several ecological factors, while not directly causing evolvability, influence the degree to which it becomes evident within a population. These factors encompass environmental stimuli, interactions amongst individuals within the same population, ecological dynamics between individuals in different populations, and varying selection pressures. Their interplay results in the creation of an adaptive landscape, which the population traverses in its quest for a global (or local) optimum. Thus far, only a limited number of theoretical works have addressed these adaptive dynamics while considering the concept of evolvability \cite{bukkuri2023, crombach2008,cuypers2017,hickinbotham2021}.

Our aim is to carry out a theoretical study of the effect of evolvability on the adaptive dynamics of phenotypically-structured cell populations. Although we will focus on the case of cells, the model is general enough to consider any population of autonomous self-replicating agents subject to evolution, such as multicellular organisms. We will use a stochastic individual-based (IB) model and corresponding deterministic continuum model, which we derive from the IB model using formal limiting procedures. The IB model is able to capture the stochastic nature of the phenotypic changes that may occur within individuals of the same population, hence recapitulating phenomena observed in the presence of small numbers of individuals such as genetic drift, while the continuum model provides a concise depiction of the collective behaviour of large numbers of individuals, thus making it possible to explore the adaptive trajectories of a whole population.

In this paper, we build upon earlier studies that addressed the evolutionary dynamics of phenotypically-structured populations \cite{chisholm2016,lorenzi2016,ardaseva2020,ardaseva2020b,ardaseva2020c} and non-genetic stochastic changes in cellular traits \cite{ortega2022}. First of all, we expand the notion of populations with fluctuations in the characteristics considered in \cite{ardaseva2020}, so that we include evolvability as a continuous phenotypic trait. In the context of the model, evolvability is defined as the degree of variability in proliferative potential, such that an individual with higher evolvability will be more likely to undergo phenotypic changes in their proliferative potential than one with lower evolvability. Taking this information into consideration, and following the idea presented in \cite{ortega2022}, since evolvability is considered as a phenotypic trait \textit{per se}, it is also subject to spontaneous heritable changes. Hence, we let the evolvability of a cell be itself subject to evolutionary change. In this way, we can explore the effect of non-genetic heterogeneity on the evolutionary dynamics of a phenotypically diverse population subject to varying adaptive potential, considering also fitness costs and selection pressures. 

\section{Methods}

\subsection{The IB model}
\label{Sec:IBModel}

We developed an IB model for the evolutionary dynamics of a phenotypically heterogeneous population consisting of \textit{cells}. We will use the term \textit{cells} from now on, although this definition can be extended to more general agents, so that agent can mean any self-replicating individual. The phenotypic state of every cell at time $t \in [0, t_f]$ is characterised by the structuring variables $y \in [0,1]$ and $x \in [0,1]$, which take into account intercellular variation in proliferative potential and evolvability, respectively. Without loss of generality, we focus on the case where larger values of proliferative potential $y$ correspond to a higher cell division rate, and larger values of evolvability $x$ correspond to a higher probability for changes in proliferative potential to occur. The proliferative potential could be represented by the normalised level of expression of a gene that regulates cell division -- such as \textit{MKI67, BIRC5, CCNB1, CDC20, CEP55, NDC80, TYMS, NUF2, UBE2C, PTTG1,} and \textit{RRM2} \cite{feitelson2015,guo2020,nielsen2010}, while the evolvability could be related to the degree of variation of the level of expression of such a gene over time and, therefore, could be connected with the normalised level of expression of a gene that controls the expression of genes regulating cell division -- such as  \textit{FOXM1}, \textit{MYBL2} or \textit{TOP2A} \cite{guo2020,ahmed2019}.

In order to define an on-lattice model, we discretise the time and phenotype variables. The current time-step $t_h$ indicates how many steps $h$ of size $\Delta t$ have already been taken. Cells will be characterised by their phenotypic state ($y_i,x_j$), which is represented as a position on the lattice $\{ y_i \}_i \times \{x_j\}_j$ corresponding to phenotypic space $[0,1] \times [0,1]$. Cells can change their proliferative potential by taking $i$ steps of size $\Delta y$ (only one per time-step), as well as their evolvability, by taking $j$ steps of size $\Delta x$ (also, only one per time-step). 

We introduce the variable $\mathcal{N}_{i,j}^h$ to model the number of cells in the phenotypic state ($y_i,x_j$) at time $t_h$. From $\mathcal{N}_{i,j}^h$, we can calculate the cell population density $n_{i,j}^{h}$ (i.e. the phenotypic distribution of the cells) and the corresponding cell population size $N^{h}$ (i.e. the total number of cells) at time $t_h$. To keep track of the evolution of the phenotypic state, we define the mean level of proliferative potential $\bar{y}^h$ and the mean level of evolvability $\bar{x}^h$ (as well as their corresponding standard deviations $\sigma_y^h$ and $\sigma_x^h$). Hence, the point $(\bar{y}^h,\bar{x}^h)$ represents the mean phenotypic state of the cell population at time $t^h$. 

\begin{figure}[h!]
\centering
\includegraphics[width=\textwidth]{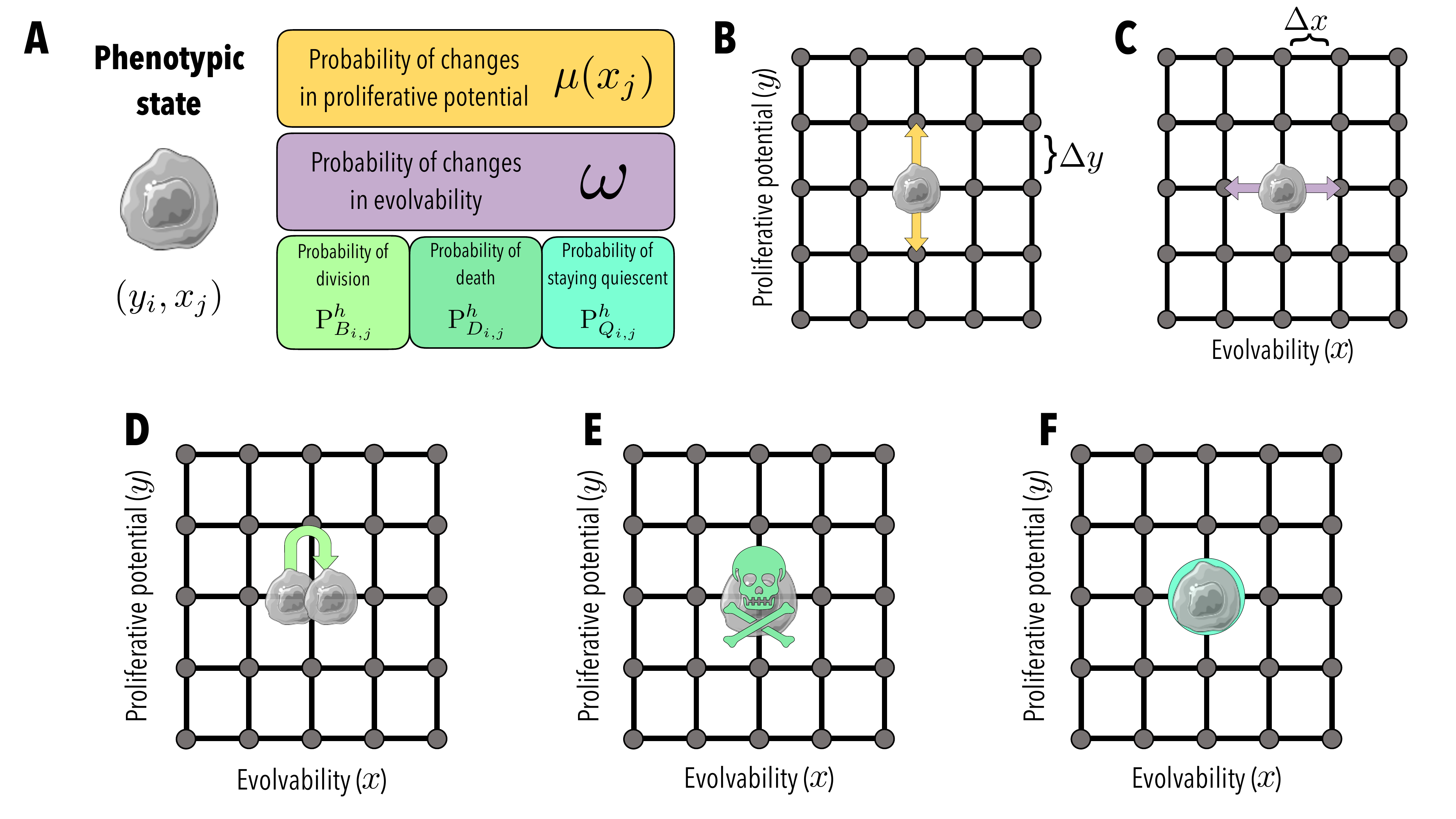}
\caption[IB model schematic]{ Schematic depiction of the basic processes a cell can undergo in the IB model. \textbf{A} A cell is characterised by its phenotypic state $(y_i,x_j)$, a pair of values of proliferative potential $y_j$ and evolvability $x_i$ that determine a position in the lattice $\{y_i\}_{i} \times \{x_j\}_{j}$, which defines the phenotypic space. Each phenotypic state $(y_i,x_j)$ is associated with a division probability $P_{B_i,j}^h$ and a probability of change in proliferative potential $\mu(x_j)$. The different processes that a cell may undergo between time-steps $h$ and $h+1$ are summarized in the color boxes, and detailed hereafter. \textbf{B} Cells may undergo spontaneous phenotypic changes that increase or decrease their proliferative potential, according to a probability $\mu(x_i)$ that depends on their level of evolvability. \textbf{C} Cells may also undergo spontaneous phenotypic changes that increase or decrease their level of evolvability, according to a fixed probability $\omega$. Increasing or decreasing the proliferative potential and the level of evolvability is equiprobable. \textbf{D} Cells may divide according to a probability ${\rm P}_{B_{i,j}}^{h}$ (cf. Eq. \eqref{Prob_birth}). The newborn cell inherits the same phenotypic state as its parent. \textbf{E} Cells may die with a probability ${\rm P}_{D_{i,j}}^{h}$ (cf. Eq. \eqref{Prob_death}). Both division and death probabilities depend on a cell's levels of proliferative potential and evolvability. \textbf{F} Cells may remain quiescent with a probability ${\rm P}_{Q_{i,j}}^{h}$ (cf. Eq. \eqref{Prob_quiescent}).}
\label{Fig1}
\end{figure} 

Time and phenotypic variables are discretised according to 
\begin{align*}
&t_h = h\Delta t \in [0,t_f], \; y_i = i \Delta y \in [0,1], \; x_j = j \Delta x \in [0,1], \\
&h,i,j \in \mathbb{Z}^+, \; \Delta t, \Delta y, \Delta x \in \mathbb{R}^{+}_*,
\end{align*}
where $\mathbb{Z}^+$ denotes the set of nonnegative integers and $\mathbb{R}^{+}_*$ denotes the set of positive real numbers -- i.e. $\mathbb{R}^{+}_* = \mathbb{R}^{+} \setminus \{0\}$. The cell population density and the corresponding cell population size are computed via the following formulas
\begin{equation}
 n_{i,j}^{h} \equiv n(t_h,y_i,x_j) := \frac{\mathcal{N}_{i,j}^h}{\Delta y \Delta x} \quad \text{and} \quad N^{h} \equiv N(t_h) := \sum_{i,j} \mathcal{N}_{i,j}^h,
 \label{eq:density_define}
\end{equation}
while the mean level of proliferative potential and the mean level of evolvability of the cell population at time $t_h$, along with the corresponding standard deviations, are computed according to
\begin{equation}
\bar{y}(t_h) \equiv \bar{y}^h := \frac{1 }{N^h} \sum_{i,j} \mathcal{N}_{i,j}^h  y_i, \quad \sigma_y(t_h) \equiv \sigma_y^h := \left(  \frac{1}{N^h} \sum_{i,j}\mathcal{N}_{i,j}^h  y_i^2 - \left(\bar{y}^h \right)^2  \right)^{1/2}
\label{eq:meanvariancey_define}
\end{equation}
and
\begin{equation}
\bar{x}(t_h) \equiv \bar{x}^h := \frac{1}{N^h} \sum_{i,j} \mathcal{N}_{i,j}^h  x_j, \quad \sigma_x(t_h) \equiv \sigma_x^h := \left(\frac{1}{N^h} \sum_{i,j}\mathcal{N}_{i,j}^h  x_j^2 - \left(\bar{x}^h \right)^2  \right)^{1/2}.
\label{eq:meanvariancex_define}
\end{equation}

As summarised by the schematics in Figure \ref{Fig1}, between time-steps $h$ and $h+1$, each cell in phenotypic state $(y_i, x_j)$ can first undergo heritable, spontaneous phenotypic changes and then die, divide, or stay quiescent according to the rules described in the following subsections.

\subsubsection{Mathematical modelling of cell division and death}

We assume that a dividing cell is instantly replaced by two identical cells that inherit the phenotypic state of the parent cell (i.e. the progenies are placed on the same lattice site as their parent), while a dying cell is instantly removed from the population. We model saturating growth of the cell population by letting the cells divide, die or remain quiescent with probabilities that depend on their phenotypic state and the cell population size. In particular, to define the probabilities of cell division and death, we introduce the function $R \equiv R(y_i, x_j, N^h)$, which describes the net division rate (i.e. the difference between the rate of division and the rate of death) of cells in the phenotypic state $(y_i, x_j)$ under the environmental conditions corresponding to the cell population size $N^h$. In particular, we will focus on the case where:

\begin{equation}
 \label{def:R}
R(y,x,N) := \rho(x,y) - \kappa \, N.
\end{equation}

The definition given by Eq.~\eqref{def:R} relies on the following assumptions: cells die due to intra-population competition at a rate proportional to the size of the cell population, with constant of proportionality $\kappa>0$; and cells in the phenotypic state $(y,x)$ divide and die due to natural selection on the proliferative potential at rate $\rho(x,y)$ (i.e. $\rho(x,y)$ is the intrinsic net division rate of cells in the phenotypic state $(y,x)$). In the framework of our model, under the definition given by Eq.~\eqref{def:R}, the function $\rho(x,y)$ determines the shape of the phenotypic fitness landscape of the cell population. 

In order to capture the fact that a larger proliferative potential corresponds to a higher cell division rate (i.e. a larger fitness), and to consider scenarios in which, due to possible fitness costs associated with evolvability~\cite{geller2018,dewitt1998,bloom2007}, cells may undergo cell division at different rates depending on their evolvability level, we define the intrinsic net division rate $\rho$ as (building on the ideas presented in ~\cite{ardaseva2020,lorenzi2016}):
\begin{equation}
\label{def:rho}
\rho(x,y) :=  \gamma \left(1-\alpha \, r(x)\right) - \eta (1-y)^2.
\end{equation}
Here, the parameter $\gamma>0$ is the maximum cell division rate (i.e. the maximum fitness) and the parameter $\eta > 0$ is a selection gradient that provides a measure of the strength of natural selection on the proliferative potential. Furthermore, the parameter $\alpha \geq 0 $ models the fitness cost of evolvability and the function $r(x) \geq 0$ models how the level of evolvability impacts on the intrinsic net division rate. To consider the scenario where higher evolvability levels decrease the rate of cell division, we define the function $r$ in Eq.\eqref{def:rho} as:
\begin{equation}
\label{def:r1}
r(x) := x^2.
\end{equation}
 Conversely, to consider the opposite scenario where cells need to invest energy in keeping low levels of evolvability (e.g. to ensure genome fidelity), and thus lower evolvability levels lower the rate of cell division, we use the following definition: 
\begin{equation}
\label{def:r2}
r(x) := (1-x)^2.
\end{equation}
The definition given by Eq.~\eqref{def:rho} is such that under scenarios in which evolvability does not imply a fitness cost (i.e. when $\alpha = 0$), cells with the highest proliferative potential (i.e. the fittest phenotypic variants with $y=1$) will divide at the maximum rate, $\gamma$, and the net division rate of cells with a lower proliferative potential (i.e. less fit phenotypic variants with $y \in [0,1)$) decreases as the selection gradient increases (i.e. larger values of $\eta$ and smaller values of $y$ correspond to a lower net division rate). Moreover, under scenarios in which there is a fitness cost associated with evolvability (i.e. when $\alpha > 0$), the net division rate of the cells decreases with the fitness cost of evolvability, $\alpha$, and higher (resp. lower) levels of evolvability correspond to a lower net division rate under the definition of $r(x)$ provided by Eq.~\eqref{def:r1} (resp. Eq.~\eqref{def:r2}). Note that, under these definitions, there may be phenotypic states $(y,x)$ for which the intrinsic net division rate is negative, implying that the rate at which cells in these phenotypic states divide is lower than the rate at which they die due to natural selection on the proliferative potential.

Using the definitions in Eqs.~\eqref{def:R} and \eqref{def:rho}, we assume that between time-steps $h$ and $h+1$ a cell in phenotypic state $(y_i, x_j)$ may divide with probability

\begin{equation}
{\rm P}_{B_{i,j}}^{h} := \Delta t \rho_+(x,y) \;\text{ where } \; \rho_+(x,y) = \max \left(0, \rho(x,y) \right),
\label{Prob_birth}
\end{equation}

die with probability

\begin{equation}
{\rm P}_{D_{i,j}}^{h} := \Delta t \left[ \kappa \, N + \rho_-(x,y) \right] \;
\text{ where } \; \rho_-(x,y) = -\min \left(0, \rho(x,y) \right),
\label{Prob_death}
\end{equation}

or remain quiescent (i.e. do not divide nor die) with probability
\begin{equation}
{\rm P}_{Q_{i,j}}^{h} := 1-{\rm P}_{B_{i,j}}^{h}-{\rm P}_{D_{i,j}}^{h}.
\label{Prob_quiescent}
\end{equation}

Note that we are implicitly assuming that the time-step size $\Delta t$ is sufficiently small so that $0 \leq {\rm P}_{B_{i,j}}^{h} + {\rm P}_{D_{i,j}}^{h} \leq 1$ for all values of $i$, $j$, and $h$. Note also that we define the positive part $\rho_+(x,y)$ and the negative part $\rho_-(x,y)$ of the division rate $\rho(x,y)$, such that $\rho_+(x,y)$ contributes to the division probability ${\rm P}_{B_{i,j}}^{h}$, meanwhile $\rho_-(x,y)$ contributes to the death probability ${\rm P}_{D_{i,j}}^{h}$.

\subsubsection{Mathematical modelling of phenotypic changes}
\label{subsec:mmpc}
We take into account phenotypic changes by allowing cells to update their phenotypic state according to a random walk along the two phenotypic dimensions. We assume that changes in the level of evolvability occur with a constant probability, whilst changes in the proliferative potential occur with a probability that increases with the evolvability level. We model these probabilities through the parameter $\omega \in [0,1]$ and the function $0 \leq \mu(x) \leq 1$, respectively, and we focus on the case where 
\begin{equation}
 \label{def:mu}
\mu(x) := x^2. 
\end{equation} 

In such a modelling framework, between time-steps $h$ and $h+1$, every cell in phenotypic state $(y_i,x_j)$ can: undergo a change in its level of evolvability, with probability $\omega$, or retain its current level of evolvability, with probability $1 - \omega$; undergo a change in its proliferative potential, with probability $\mu(x_j)$, or retain its current proliferative potential, with probability $1 - \mu(x_j)$. We assume that a cell in phenotypic state $(y_i,x_j)$ that undergoes a change in its level of evolvability acquires either of the evolvability levels corresponding to $x_{j\pm1} = x_{j} \pm  \Delta x$ with probabilities $\omega/2$. Conversely, building on~\cite{chisholm2016discrete}, we assume that a cell in the phenotypic state $(y_i,x_j)$ that undergoes a change in its proliferative potential may acquire a lower level of proliferative potential ($y_{i-1} = y_{i} -  \Delta y$) with probability $\theta \in [0,1]$, or a higher level of proliferative potential ($y_{i+1} = y_{i} +  \Delta y$) with probability $1-\theta$. When $\theta = 0.5$, there is a symmetrical distribution of fitness effects (DFE), meaning that cells are equally likely to acquire a higher or lower proliferative potential (i.e. phenotypic changes are equally likely to be advantageous or deleterious). First we will study this scenario. Then, since detrimental phenotypic changes may be more likely than beneficial ones \cite{eyre2007}, we will also explore scenarios in which $\theta > 0.5$, whereby there is an asymmetrical DFE and cells are more likely to acquire a lower proliferative potential as a result of phenotypic changes. Any attempted phenotypic variation of a cell that would require moving into a state outside the phenotypic domain $[0,1] \times [0,1]$ is aborted.
	
\subsection{The corresponding continuum model}
\label{Sec:PIDEModel}
Through a method analogous to that we previously employed in~\cite{ardaseva2020b,bubba2020discrete,chaplain2020bridging,chisholm2016discrete}, letting the time-step size $\Delta t\rightarrow 0^+$ and the sizes of phenotype-steps $\Delta y \rightarrow0^+$ and $\Delta x \rightarrow0^+$, as well as $\theta \rightarrow0.5^+$, in such a way that
\begin{equation}
\frac{\left(\Delta y\right)^2}{2\Delta t}  \rightarrow \xi_1 \in \mathbb{R}^+_*, \quad \quad \frac{\left(\Delta x\right)^2}{2\Delta t}  \rightarrow \xi_2 \in \mathbb{R}^+_*, \quad \text{and} \quad  \frac{(2\theta -1)\Delta y}{\Delta t}  \rightarrow \xi_3 \in \mathbb{R},
\label{derived_parameters}
\end{equation}
we formally show (see the Appendix) that the deterministic continuum counterpart of the stochastic IB model presented in the previous section is given by the following partial integro-differential equation (PIDE) for the cell population density function $n(t,y,x) \geq 0$:
\begin{equation}
\label{eq:PIDE}
\begin{cases}
&\partial_t n = R(y,x,N) \, n + \xi_1 \mu(x) \, \partial_y \left(\partial_y n + \dfrac{\xi_3}{\xi_1} n \right) + \xi_2\omega \, \partial^2_{xx} n, \\
&\phantom{\partial_t n = R} (t,y,x) \in (0,t_f] \times (0,1) \times (0,1),
\\ \\
&\displaystyle{N(t) := \int_{0}^1 \int_{0}^1 n(t,y,x) \, \textrm{d}y \,\textrm{d}x},
\end{cases}
\end{equation}
subject to zero Neumann (i.e. no-flux) boundary conditions on the boundary of the square $[0,1] \times [0,1]$. In the continuum modelling framework given by the PIDE~\eqref{eq:PIDE}, the function $n(t,y,x)$ represents the number density of cells in the phenotypic state $(y,x)$ and its integral, $N(t)$, is thus the size of the cell population at time $t$. The reaction term models the effects of cell division and death, and the function $R(y,x,N)$, which is defined via Eq.~\eqref{def:R}, represents the net division rate of cells in the phenotypic state $(y,x)$ under the environmental conditions corresponding to the cell population size $N$. Moreover, the diffusion term in the $x-$direction models the effects of changes in the level of evolvability, which occur at rate $\xi_2\omega$. Finally, the diffusion-advection term in the $y-$direction models the effects of changes in the level of proliferative potential, which occur at rate $\xi_1 \mu(x)$. The advection term captures the possible presence of asymmetry in the DFE (cf. Subsection~\ref{subsec:mmpc}), and the ratio $\dfrac{\xi_3}{\xi_1}$ measures the relative impact of asymmetrical DFE on cell dynamics. In the case of symmetrical DFE, $\theta=0.5$ in the underlying IB model and thus $\xi_3=0$. On the other hand, in the case of asymmetrical DFE, if phenotypic changes are more likely to have negative effects, meaning that cells are more likely to acquire a lower level of proliferative potential, then $\theta>0.5$, thereby $\xi_3>0$.

The mean levels of proliferative potential \eqref{eq:meanvariancey_define} and evolvability \eqref{eq:meanvariancex_define} defined in Section~\ref{Sec:IBModel} have the following corresponding functions in the continuum counterpart:

\begin{equation}
\begin{split}
\bar{y}(t) &:= \frac{1}{N(t)} \int_{0}^1 \int_{0}^1 y \, n(t,y,x) \, \textrm{d}y \,\textrm{d}x, \\
\sigma_y(t) &:= \left(\frac{1}{N(t)} \int_{0}^1 \int_{0}^1 y^2 \, n(t,y,x) \, \textrm{d}y \,\textrm{d}x - \left(\bar{y}(t) \right)^2  \right)^{1/2}
\end{split}
\label{eq:meanvarianceyPIDE_define}
\end{equation}
and
\begin{equation}
\begin{split}
\bar{x}(t) &:= \frac{1}{N(t)} \int_{0}^1 \int_{0}^1 x \, n(t,y,x) \, \textrm{d}x \,\textrm{d}y, \\
\sigma_x(t) &:= \left(\frac{1}{N(t)} \int_{0}^1 \int_{0}^1 x^2 \, n(t,y,x) \, \textrm{d}x \,\textrm{d}y - \left(\bar{x}(t) \right)^2  \right)^{1/2}.
\end{split}
\label{eq:meanvariancexPIDE_define}
\end{equation}

\section{Main results}\label{sec2g}
In this section, we present the main results of numerical simulations of the IB model, which we compare with numerical solutions of the continuum model given by the PIDE~\eqref{eq:PIDE}. The set-up of numerical simulations is summarised in Subsection~\ref{sec:setupnum}.

In Subsection~\ref{sec:ressymDFE}, focussing on the scenario of symmetrical DFE (i.e. when the IB model parameter $\theta=0.5$ and, therefore, the PIDE model parameter $\xi_3=0$), we first consider the case where evolvability does not imply a fitness cost (i.e. $\alpha = 0$ in Eq.~\eqref{def:rho}) and present a sample of base-case results that summarise the evolutionary dynamics of the cell population for different choices of the initial mean level of proliferative potential and the initial mean level of evolvability. We then present the results of numerical simulations carried out to investigate how the base-case evolutionary dynamics change as we vary the values of the parameters $\eta$ and $\alpha$ in the definition given by Eq.~\eqref{def:rho}, in order to explore how the strength of natural selection on the proliferative potential, the fitness cost of evolvability (given by $\alpha > 0$ in Eq.~\eqref{def:rho}), and the interplay between these evolutionary parameters affect the phenotypic evolution of the cell population. In particular, we consider situations where higher evolvability levels decrease the rate of cell division~\cite{snell2010,giraud2001}, and thus complement Eq.~\eqref{def:rho} with Eq.~\eqref{def:r1}. Although this ideal set-up may be far away from reality, it provides valuable insight into the model dynamics.

In Subsection~\ref{sec:resasymDFE}, we turn to the scenario of asymmetrical DFE whereby phenotypic changes are more likely to make cells acquire a lower proliferative potential (i.e. when the IB model parameter $\theta>0.5$, and thus the PIDE model parameter $\xi_3>0$). Since in this scenario an `endogenous' fitness cost is already placed on higher levels of evolvability by asymmetrical DFE, we first set $\alpha = 0$ in Eq.~\eqref{def:rho}. We then study the interplay between such an endogenous cost of evolvability and an `exogenous' cost, meaning that cells need to invest energy to maintain low levels of evolvability. To this end, we keep $\theta > 0.5$ (i.e. $\xi_3>0$) and set $\alpha > 0$ in Eq.~\eqref{def:rho} complemented with Eq.~\eqref{def:r2}.

Generally, under the choices made here for the parameter values, the simulation results demonstrate excellent quantitative agreement between the stochastic IB model and its deterministic continuum counterpart given by the PIDE \eqref{eq:PIDE}. This testifies to the robustness of these results, and thus of the possible biological conclusions drawn therefrom (nonetheless limited by the assumptions made in the definition of the IB model). 

\subsection{Set-up of numerical simulations}
\label{sec:setupnum}
For consistency with previous mathematical studies of the evolutionary dynamics of phenotype-structured populations, which rely on the \textit{prima facie} assumption that population densities are Gaussians~\cite{rice2004evolutionary}, simulations are carried out under the assumption that the initial phenotype distribution of cells for the IB model is defined as:
\begin{equation} \label{eq:ICIB}
n^0_{i,j} := \frac{\mathcal{N}_{i,j}^0}{\Delta y \Delta x} , \quad \mathcal{N}_{i,j}^0 := N^0 \, C \, \exp\left[-\frac{(y_i-\bar{y}^0)^2}{2(\sigma^0_y)^2}  - \frac{(x_j-\bar{x}^0)^2}{2(\sigma^0_x)^2}\right] ,
\end{equation}
where $C$ is a normalisation constant such that $\displaystyle{\sum_{i,j} \mathcal{N}^0_{i,j} = N^0}$. In the definition given by Eq.~\eqref{eq:ICIB}, the parameter $N^0$ represents the initial cell number, the parameters $\bar{y}^0$ and $\sigma^0_y$ represent the initial mean proliferative potential and the corresponding standard deviation, respectively, while the parameters $\bar{x}^0$ and $\sigma^0_x$ represent the initial mean level of evolvability and the corresponding standard deviation, respectively. The (adimensional) baseline parameter values used to carry out numerical simulations of the IB model are listed in Table~\ref{Table1}. Note that, consistently with the conditions given by Eq.~\eqref{derived_parameters}, the values of $\Delta t$, $\Delta y$, and $\Delta x$ are chosen sufficiently close to $0$, while the values of the parameter $\theta$ are chosen sufficiently close to $0.5$. The methods employed to numerically solve the PIDE~\eqref{eq:PIDE} subject to no-flux boundary conditions and to an initial condition $n(0,y,x)$, which is the continuum analogue of $n^0_{i,j}$ defined via Eq.~\eqref{eq:ICIB}, are described in the Appendix.

\setlength{\extrarowheight}{6pt}
\setlength{\tabcolsep}{6pt}
\begin{table*}[hbtp]
\caption{Baseline parameter values used to carry out the  numerical simulations.}
\label{Table1}
%{\rowcolors{2}{green!80!yellow!50}{green!70!yellow!40}
{\rowcolors{2}{blue!10!}{blue!20!}
\begin{tabular}{|c|c|c|}
\hline
\textbf{Parameter} & \textbf{Biological Meaning} & \textbf{Value} \\ \hline
{$t_{f}$} & \footnotesize{Final time} & {$\{5\times 10^2,10^3\}$} \\ \hline
{$\Delta t$} & \footnotesize{Time-step size}  & {$10^{-4}$} \\ \hline
{$\Delta y$, $\Delta x$} & \footnotesize{Phenotype-step size}  & {$\approx \, 0.0141$} \\ \hline
{$N_0$} & \footnotesize{Initial cell number} & {$100$} \\ \hline
{$\bar{y}^0$} & \footnotesize{Initial mean proliferative potential} & { $\{0.2,0.6,0.8\}$ } \\ \hline
{$\sigma^0_y$} & \footnotesize{Standard deviation corresponding to $\bar{y}^0$} & {$0.02$} \\ \hline
{$\bar{x}^0$} & \footnotesize{Initial mean level of evolvability} & {$\{0.2,0.7,0.8\}$} \\ \hline
{$\sigma^0_x$} & \footnotesize{Standard deviation corresponding to $\bar{x}^0$} & {$0.02$} \\ \hline
{$\omega$} & \footnotesize{Probability of changes in evolvability level} & {$4 \times 10^{-3}$}  \\ \hline
{$\gamma$} & \footnotesize{Maximum fitness} & {$1$} \\ \hline
{$\eta$} & \footnotesize{Gradient of selection on proliferative potential} & {$\{ 0.05,0.2,0.5,0.8,1.5 \}$ }  \\ \hline
{$\alpha$} & \footnotesize{Fitness cost of evolvability} & {$\{ 0.125,0.25,0.5,1,1.5 \}$}  \\ \hline
{$\kappa$} & \footnotesize{Rate of cell death due to intra-population competition} & {$10^{-4}$}  \\ \hline
$\theta$ & \footnotesize{Probability of acquiring a lower level of proliferative potential} & \{$0.5$, $0.55$, $0.6$, $0.7$\}  \\ \hline
\end{tabular}}
\end{table*}

\newpage
\subsection{Main results under symmetrical DFE}
\label{sec:ressymDFE}

We start by considering symmetrical DFE whereby phenotypic changes are equally likely to be advantageous or deleterious (i.e. to make cells acquire a higher or lower proliferative potential). Hence, we investigate scenarios where the IB model parameter $\theta = 0.5$, and thus the PIDE model parameter $\xi_3=0$.

\subsubsection{{\normalsize Different initial phenotypic compositions lead to the same evolutionary trajectory}}
\label{Sec:basecaseresults}

In a realistic setting, different cell populations would be expected to have different phenotypic compositions, depending on the level of adaptation to their environment. We investigated how the initial phenotypic distribution of a population can affect its evolutionary dynamics. In particular, we considered four different scenarios, whereby the initial phenotypic composition of a cell population corresponds to  the possible combinations between low/high evolvability and low/high proliferative potential. Then, we simulated the evolutionary trajectories of these populations using the IB and the PIDE models.

\begin{figure}[h!]
\centering
\includegraphics[width=1\textwidth]{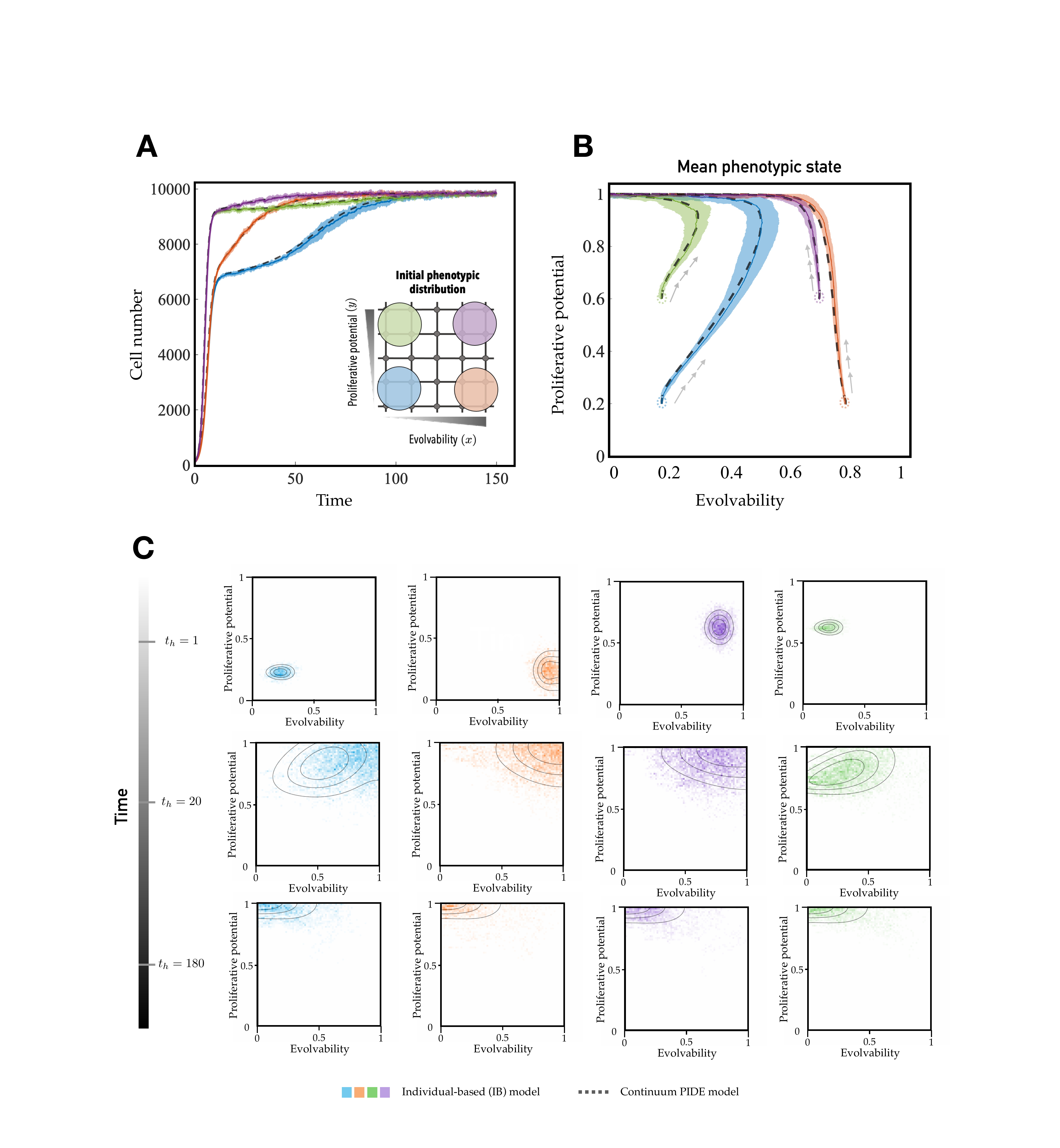}
\caption{{\bf Influence of the initial phenotypic composition on the evolutionary dynamics under symmetrical DFE.} \textbf{A}-\textbf{B} Dynamics of the cell number (panel \textbf{A}) and the mean phenotypic state (panel \textbf{B}). Solid coloured lines display the results of numerical simulations of the IB model while dashed black lines display the results of numerical simulations of the continuum model, when evolvability does not imply a fitness cost (i.e. $\alpha = 0$), for different values of the initial mean levels of evolvability and proliferative potential -- i.e. $(\bar{x}^0, \bar{y}^0)=(0.2,0.2)$ (blue lines), $(\bar{x}^0, \bar{y}^0)=(0.2,0.6)$ (green lines), $(\bar{x}^0, \bar{y}^0)=(0.8,0.2)$ (orange lines), and $(\bar{x}^0, \bar{y}^0)=(0.7,0.6)$ (purple lines). The results from the IB model correspond to the average over 20 simulations and the related standard deviation is displayed by the coloured areas surrounding the curves. \textbf{C} Phenotypic distribution of the cell population at three different time points for each initial scenario considered. Contour lines represent areas with the same cell density in the numerical simulations of the continuum model, while coloured lattice points in the phenotypic domain represent cell density at each phenotypic state (where a greater colour intensity indicates a higher cell density). Numerical simulations of the IB model were carried out using the initial phenotypic distribution defined via Eq.~\eqref{eq:ICIB} and the parameter values listed in Table~\ref{Table1} with $\alpha=0$ and $\eta=0.5$ in Eq.~\eqref{def:rho}. Details of numerical simulations of the continuum model are provided in the Appendix.} 
\label{Fig2}
\end{figure}

The simulation results in Figure~\ref{Fig2}{\bf B} show the evolutionary trajectories of the cell population (i.e. the dynamics of the mean phenotypic state) under the aforementioned scenarios, corresponding to different initial mean levels of proliferative potential and evolvability, when evolvability does not imply a fitness cost (i.e. when $\alpha=0$). These results demonstrate that the mean level of proliferative potential converges asymptotically to the maximum value $1$ while the mean level of evolvability converges asymptotically to the minimum value $0$. Moreover, while the mean level of proliferative potential increases monotonically with time in all scenarios considered, the mean level of evolvability either decreases monotonically with time or first increases and then decreases depending on whether its initial value is sufficiently low or sufficiently high, respectively. In the former case, the mean level of evolvability undergoes transient growth as long as the mean level of proliferative potential is sufficiently low, and then starts decreasing monotonically, converging eventually to $0$, as soon as the mean level of proliferative potential is sufficiently high. In all scenarios, the cell phenotype distribution remains unimodal (cf. Figure~\ref{Fig2}{\bf C}). Furthermore, higher initial mean levels of proliferative potential and evolvability correlate with a faster increase in the size of the cell population, which in all scenarios saturates at a positive asymptotic value (cf. Figure~\ref{Fig2}{\bf A}).

These results suggest that for a cell population (as described by the IB model) to reach its optimal proliferative potential (and stay there), a two-step adaptive trajectory takes place. First, there is an investment in evolvability, since greater levels of this trait help in traversing the phenotypic landscape faster. Then, once the optimal proliferative potential has been reached, evolvability decreases to help cells remain at this optimal proliferative potential. The first step is skipped by cell populations whose initial level of evolvability is high enough. According to the numerical results, this adaptive trajectory seems to be robust, since all simulated cell populations underwent it.

\subsubsection{{\normalsize Stronger selection leads to faster adaptation}}

The parameter $\eta$ in Eq. \eqref{def:rho} denotes the selection gradient and thus it is a proxy for the strength of selection acting on the cell population. After computing the evolutionary trajectories that arise during the adaptation of cell populations with different initial phenotypic distributions, now we focus on the scenario where the cell population has low initial levels of both evolvability and proliferative potential, and we evaluate how different values of selection strength affect the evolutionary dynamics of the cell population.

\begin{figure}[h!]
\centering
\includegraphics[width=\textwidth]{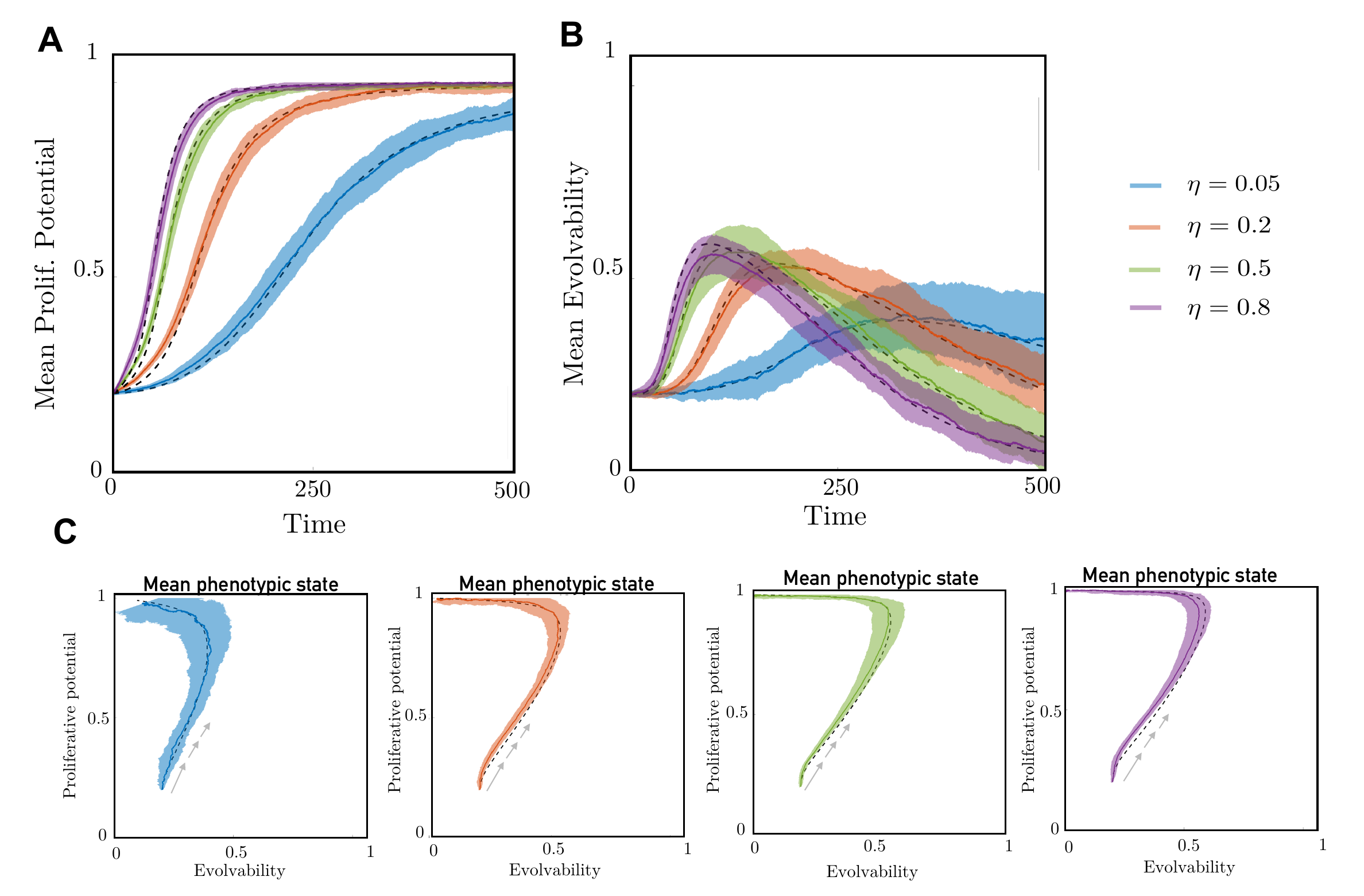}
\caption{{\bf How the strength of natural selection on the proliferative potential affects the evolutionary dynamics under symmetrical DFE.} Dynamics of the mean level of proliferative potential (panel \textbf{A}), the mean level of evolvability (panel \textbf{B}), and the mean phenotypic state (panel \textbf{C}). Solid coloured lines display the averaged results of numerical simulations of the IB model while dashed black lines display the results of numerical simulations of the continuum model, when evolvability does not imply a fitness cost (i.e. $\alpha = 0$), for different values of the selection gradient $\eta$  -- i.e. $\eta=0.05$ (blue lines), $\eta=0.2$ (orange lines), $\eta=0.5$ (green lines), and $\eta=0.8$ (purple lines). The results from the IB model correspond to the average over 20 simulations and the related standard deviation is displayed by the coloured areas surrounding the curves. Numerical simulations of the IB model were carried out using the initial phenotypic distribution defined via Eq.~\eqref{eq:ICIB} and the parameter values listed in Table~\ref{Table1} with $\alpha=0$ in Eq.~\eqref{def:rho} and $(\bar{x}^0,\bar{y}^0)=(0.2,0.2)$. Details of numerical simulations of the continuum model are provided in the Appendix. 
\label{Fig3}}
\end{figure}

The simulation results in Figure~\ref{Fig3} illustrate how the strength of natural selection on the proliferative potential (i.e. the value of the selection gradient $\eta$) affects the evolutionary dynamics of the cell population when evolvability does not imply a fitness cost (i.e. when $\alpha=0$). These results demonstrate that the larger the value of the selection gradient $\eta$, the faster are both the convergence of the mean level of proliferative potential to the maximum value $1$ and the convergence of the mean level of evolvability to the minimum value $0$. These results also indicate that in scenarios under which the mean level of evolvability undergoes transient growth, as discussed in Section~\ref{Sec:basecaseresults}, larger values of the selection gradient $\eta$ cause the mean level of evolvability of the cell population to attain larger values in the transient. For all values of $\eta$ considered here, the phenotype distribution of the cells remains unimodal and the size of the cell population saturates asymptotically to a positive value (results not displayed). 

These results suggest that the strength of selection may influence the speed of adaptation. Moreover, regardless of the value of $\eta$, all simulated populations underwent an initial increase in the level of evolvability, followed by a decrease in this level once the optimal proliferation potential was reached. Hence, we also observe that the two-step adaptive trajectory previously unveiled is robust against changes in the strength of selection. 

\subsubsection{{\normalsize What if evolvability comes at a cost?}}
\label{sec:prevsec33}

\begin{figure}[h!]
\centering
\includegraphics[width=\textwidth]{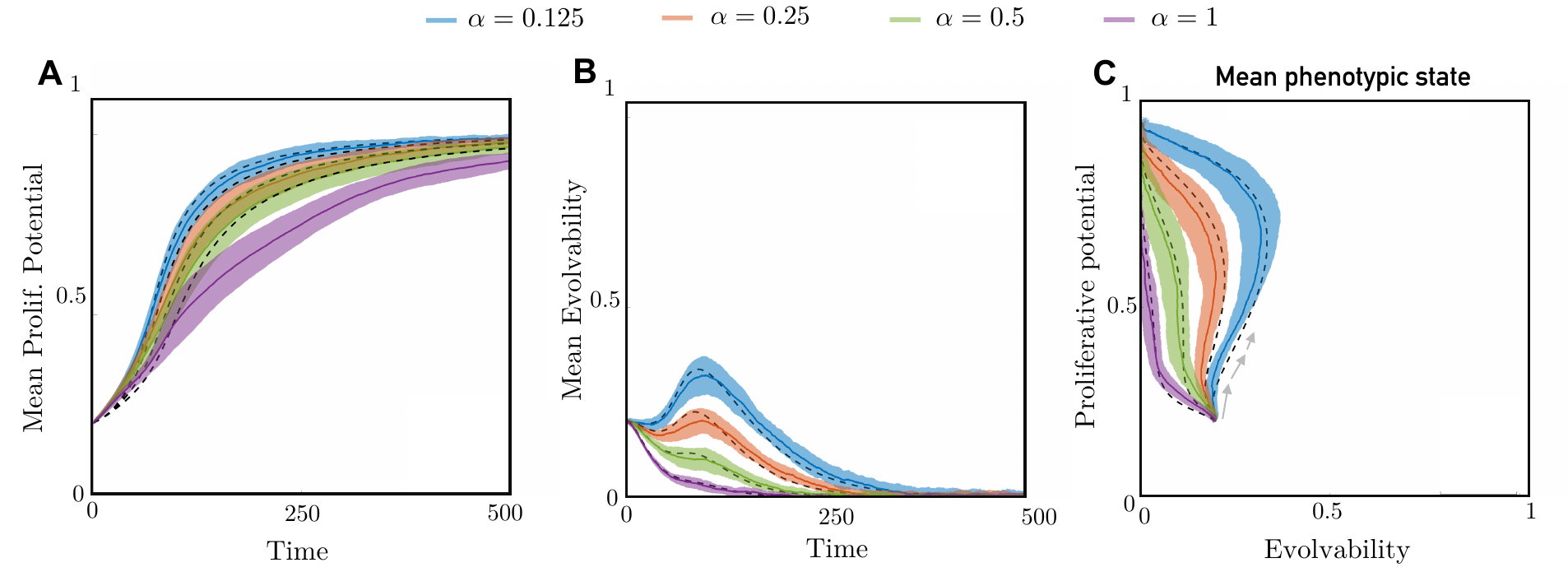}
\caption{{\bf How the fitness cost of high levels of evolvability affects the evolutionary dynamics under symmetrical DFE.} Dynamics of the mean level of proliferative potential (panel \textbf{A}), the mean level of evolvability (panel \textbf{B}), and the mean phenotypic state (panel \textbf{C}). In all panels, solid coloured lines display the results of numerical simulations of the IB model while dashed black lines display the results of numerical simulations of the continuum model, for different values of the fitness cost of evolvability $\alpha$ -- i.e. $\alpha=0.125$ (blue line), $\alpha=0.25$ (orange line), $\alpha=0.5$ (green line), and $\alpha=1$ (purple line). The results from the IB model correspond to the average over 20 simulations and the related standard deviation is displayed by the coloured areas surrounding the curves. The grey arrows in panel \textbf{C} indicate the direction of the phenotypic state trajectories. Numerical simulations of the IB model were carried out using the initial phenotypic distribution defined via Eq.~\eqref{eq:ICIB}, the definition of the intrinsic net division rate given by Eq.~\eqref{def:rho} complemented with Eq.~\eqref{def:r1}, and the parameter values listed in Table~\ref{Table1} with $\eta=0.5$ and $(\bar{x}^0,\bar{y}^0)=(0.2,0.2)$. Details of numerical simulations of the continuum model are provided in the Appendix.
\label{Fig4}
}
\end{figure}

So far, we have considered that evolvability does not imply a fitness cost. Now we turn our attention to situations where higher levels of evolvability incur a higher fitness cost \cite{snell2010,giraud2001}, and thus set $\alpha > 0$ in Eq.~\eqref{def:rho} complemented with Eq.~\eqref{def:r1}. The simulation results in Figure~\ref{Fig4} show how the evolutionary dynamics of the cell population change when there is a fitness cost associated with high levels of evolvability. These results demonstrate that larger values of the fitness cost of evolvability $\alpha$: promote a faster convergence of the mean level of evolvability to the minimum value $0$ (cf. Figure~\ref{Fig4}{\bf B} and Figure~\ref{Fig4}{\bf C}); lead to a slower convergence of the mean level of proliferative potential to the maximum value $1$ (cf. Figure~\ref{Fig4}{\bf A} and Figure~\ref{Fig4}{\bf C}); and hinder possible transient growth of the mean level of evolvability (cf. Figure~\ref{Fig4}{\bf B} and Figure~\ref{Fig4}{\bf C}). The cell phenotype distribution remains unimodal and the size of the cell population saturates at a positive asymptotic value for all values of $\alpha$ considered here (results not shown). 

These results suggest that when cells cannot excel at both phenotypic traits (i.e. proliferative potential and evolvability), the convergence to the optimal proliferation potential still happens, but not always via the two-step adaptive trajectory. In fact, the higher the fitness cost of evolvability, the less likely it is that the two-step trajectory will occur. Moreover, a higher fitness cost of evolvability also leads to a slower convergence to the optimal proliferative potential. Since cells cannot rely on greater levels of evolvability to help themselves traverse faster the phenotypic landscape, they must settle for a slower pace of adaptation.

As an aside, we observe a mismatch in the dynamics of the proliferative potential between the results of the simulations with the IB model and the numerical results of the PIDE model when the fitness cost of evolvability is relatively large (cf. Figure~\ref{Fig4}{\bf A}). This is explained by the small cell number attained by the simulations with the IB model, compared to the numerical results of the PIDE model. Since a higher cost of evolvability means a smaller division probability, it will take more time for cells in the IB model to reach the maximum value of proliferative potential (compared to the PIDE model). In such circumstances stochastic effects, not captured by the deterministic PIDE model, play a more prominent role. Hence, the cell number grows slowly and this is the reason why the mismatch is larger for greater fitness costs of evolvability.

We now investigate how the interplay between the strength of natural selection on the proliferative potential and the fitness cost of evolvability affect the evolutionary dynamics of the cell population. Since the division rate of a cell is influenced by both $\alpha$ and $\eta$ (cf. Eq. \eqref{def:rho}), taking different combinations of these parameters defines different fitness landscapes, the steepness of which will determine the ease with which a population can traverse them. Hence, we study the evolutionary dynamics of the cell population subject to different combinations of $\alpha$ and $\eta$, to define fitness landscapes with varying strength of selection and fitness cost of evolvability.

\begin{figure}[h!]
\centering
\includegraphics[width=\textwidth]{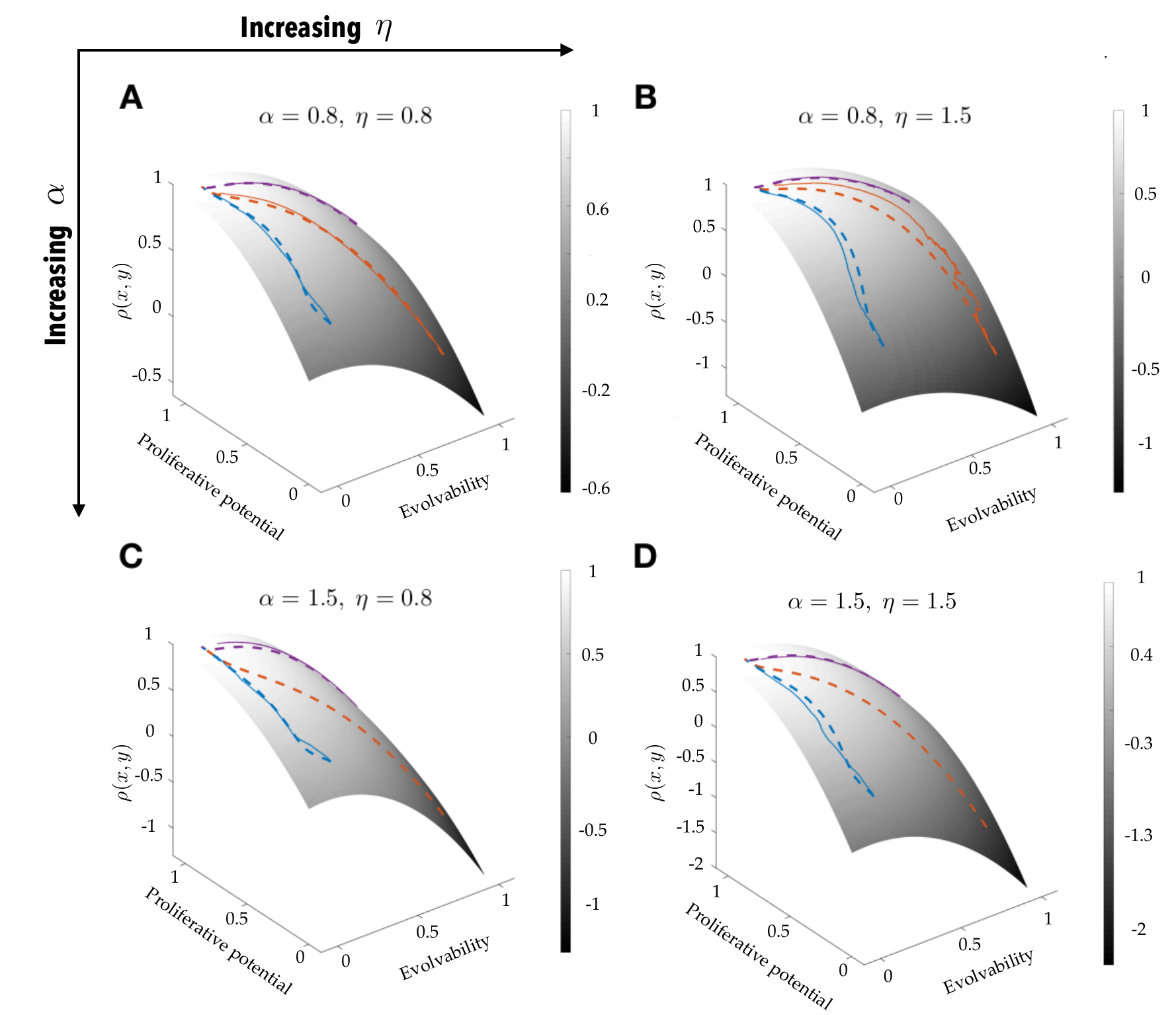}
\caption{{\bf How the interplay between the strength of natural selection on the proliferative potential and the fitness cost of evolvability affects the evolutionary dynamics under symmetrical DFE.} Dynamics of the mean phenotypic state superimposed onto the plot of the intrinsic net division rate. Results are shown for $t_{h} \in [0, 500]$. The grey-scale surfaces are the plots of the function $\rho(x,y)$ defined via Eq.~\eqref{def:rho} complemented with Eq.~\eqref{def:r1}, for the different values of $\eta$ and $\alpha$ considered. Solid coloured lines display the results of averaged numerical simulations of the IB model while dashed coloured lines display the results of numerical simulations of the continuum model, for different values of the selection gradient $\eta$ and the fitness cost of evolvability $\alpha$: $\eta=0.8$ and $\alpha=0.8$ (panel \textbf{A}), $\eta=1.5$ and $\alpha=0.8$ (panel \textbf{B}), $\eta=0.8$ and $\alpha=1.5$ (panel \textbf{C}), and $\eta=1.5$ and $\alpha=1.5$ (panel \textbf{D}), under various scenarios corresponding to different values of the initial mean levels of evolvability and proliferative potential -- i.e. $(\bar{x}^0, \bar{y}^0)=(0.2,0.2)$ (blue lines), $(\bar{x}^0, \bar{y}^0)=(0.8,0.2)$ (orange lines), and $(\bar{x}^0, \bar{y}^0)=(0.8,0.8)$ (purple lines). The results from the IB model correspond to the average over 10 simulations. Numerical simulations of the IB model were carried out using the initial phenotypic distribution defined via Eq.~\eqref{eq:ICIB} and the parameter values listed in Table~\ref{Table1}. Details of numerical simulations of the continuum model are provided in the Appendix. 
\label{Fig5}
}
\end{figure}

\begin{figure}[h!]
\centering
\includegraphics[width=\textwidth]{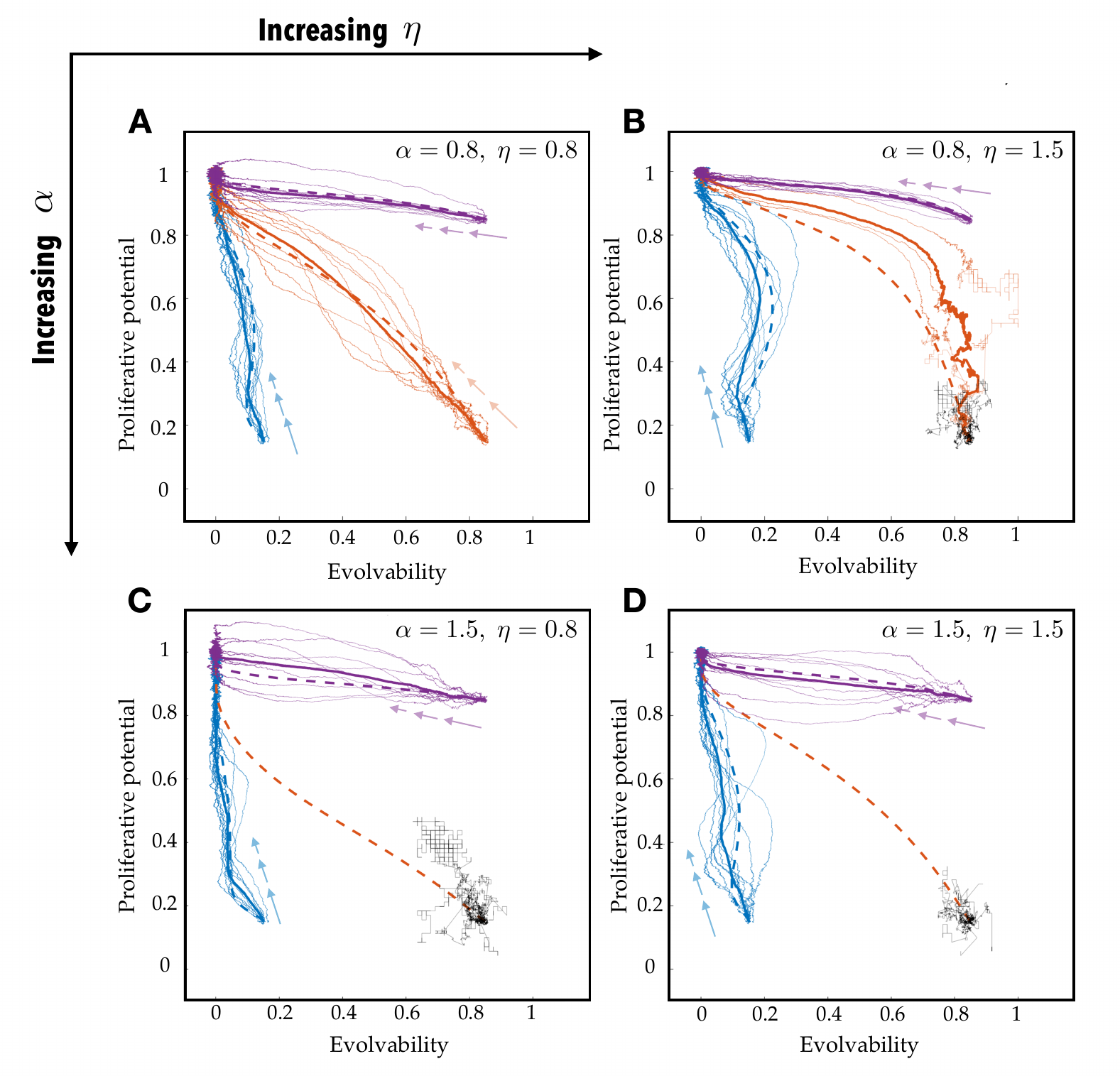}
\caption{{\bf How the interplay between the strength of natural selection on the proliferative potential and the fitness cost of evolvability affects the evolutionary dynamics under symmetrical DFE.} Dynamics of the mean phenotypic state for $t_{h} \in [0, 500]$. Solid, thin coloured lines display the results of single numerical simulations of the IB model; thick coloured lines display the average over the results of 10 numerical simulations of the IB model, and broken, coloured lines display the results of numerical simulations of the continuum model, for different values of the selection gradient $\eta$ and the fitness cost of evolvability $\alpha$ -- i.e. $\eta=0.8$ and $\alpha=0.8$ (panel \textbf{A}), $\eta=1.5$ and $\alpha=0.8$ (panel \textbf{B}), $\eta=0.8$ and $\alpha=1.5$ (panel \textbf{C}), and $\eta=1.5$ and $\alpha=1.5$ (panel \textbf{D}), under various scenarios corresponding to different values of the initial mean levels of evolvability and proliferative potential -- i.e. $(\bar{x}^0, \bar{y}^0)=(0.2,0.2)$ (blue lines), $(\bar{x}^0, \bar{y}^0)=(0.8,0.2)$ (orange lines), and $(\bar{x}^0, \bar{y}^0)=(0.8,0.8)$ (purple lines). The arrows indicate the direction of the phenotypic state trajectories. The results from the IB model correspond to the simulations displayed in Figure~\ref{Fig5}. Solid coloured lines are not displayed in the cases where the cell population goes extinct in each of the 10 simulations, with the results of simulations in which the cell population goes extinct being highlighted in black. Numerical simulations of the IB model were carried out using the initial phenotypic distribution defined via Eq.~\eqref{eq:ICIB}, the definition of the intrinsic net division rate given by Eq.~\eqref{def:rho} complemented with Eq.~\eqref{def:r1}, and the parameter values listed in Table~\ref{Table1}. Details of numerical simulations of the continuum model are provided in the Appendix. 
\label{Fig6}
}
\end{figure}

The simulation results in Figures~\ref{Fig5} and~\ref{Fig6} complement the results in Figures~\ref{Fig3} and~\ref{Fig4} by showing the impact that the interplay between the strength of natural selection on the proliferative potential and the fitness cost of evolvability may have on the evolutionary dynamics of the cell population. These results demonstrate that large values of the selection gradient $\eta$ and the fitness cost of evolvability $\alpha$ can cause extinction of cell populations in the IB model when the mean level of evolvability is initially high and the mean level of proliferative potential is initially low (cf. Figures~\ref{Fig5}{\bf C} and~\ref{Fig5}{\bf D}). This is due to the fact that sufficiently large values of these evolutionary parameters can shape the phenotypic landscape of cell populations in such a way that regions of the landscape corresponding to high levels of evolvability and low levels of proliferative potential are characterised by a negative fitness, i.e. larger values of $\eta$ and $\alpha$ may lead the intrinsic net division rate $\rho(x,y)$ in \eqref{def:rho} to attain negative values for $(x,y)$ sufficiently close to $(1,0)$, thus causing the cell population to suffer a sharp drop in its size if the initial mean phenotypic state lies in these regions (cf. Figures~\ref{Fig6}{\bf C} and~\ref{Fig6}{\bf D}). This can create the conditions for demographic stochasticity to come into play and drive the cell population to extinction (cf. Figures~\ref{Fig5}{\bf C} and~\ref{Fig5}{\bf D}). Note that, when this happens, the match between the IB model and the continuum model given by the PIDE \eqref{eq:PIDE} deteriorates, since the latter is not capable of capturing population extinction phenomena that are caused by stochastic effects associated with small cell numbers.

These results suggest that evolvability is a necessary but not sufficient condition for adaptation to occur. Under very stringent circumstances, such as the ones attained when the fitness cost of evolvability and the strength of selection are sufficiently high, a cell population may not benefit from having a high level of evolvability. Since the resources diverted to evolvability are not being used to sustain a high proliferative potential, cell populations subject to strong stochastic effects (i.e. small cell number, etc) may not succeed in their attempt to survive.

\subsection{Main results under asymmetrical DFE}
\label{sec:resasymDFE}

To investigate the impact that asymmetrical DFE, whereby phenotypic changes are more likely to be deleterious (i.e to make cells acquire a lower proliferative potential), in this subsection we explore scenarios where the IB model parameter $\theta>0.5$, and thus the PIDE model parameter $\xi_3>0$.

\subsubsection{How does asymmetrical DFE affect evolutionary dynamics?}
\begin{figure}[h!]
\centering
\includegraphics[width=\textwidth]{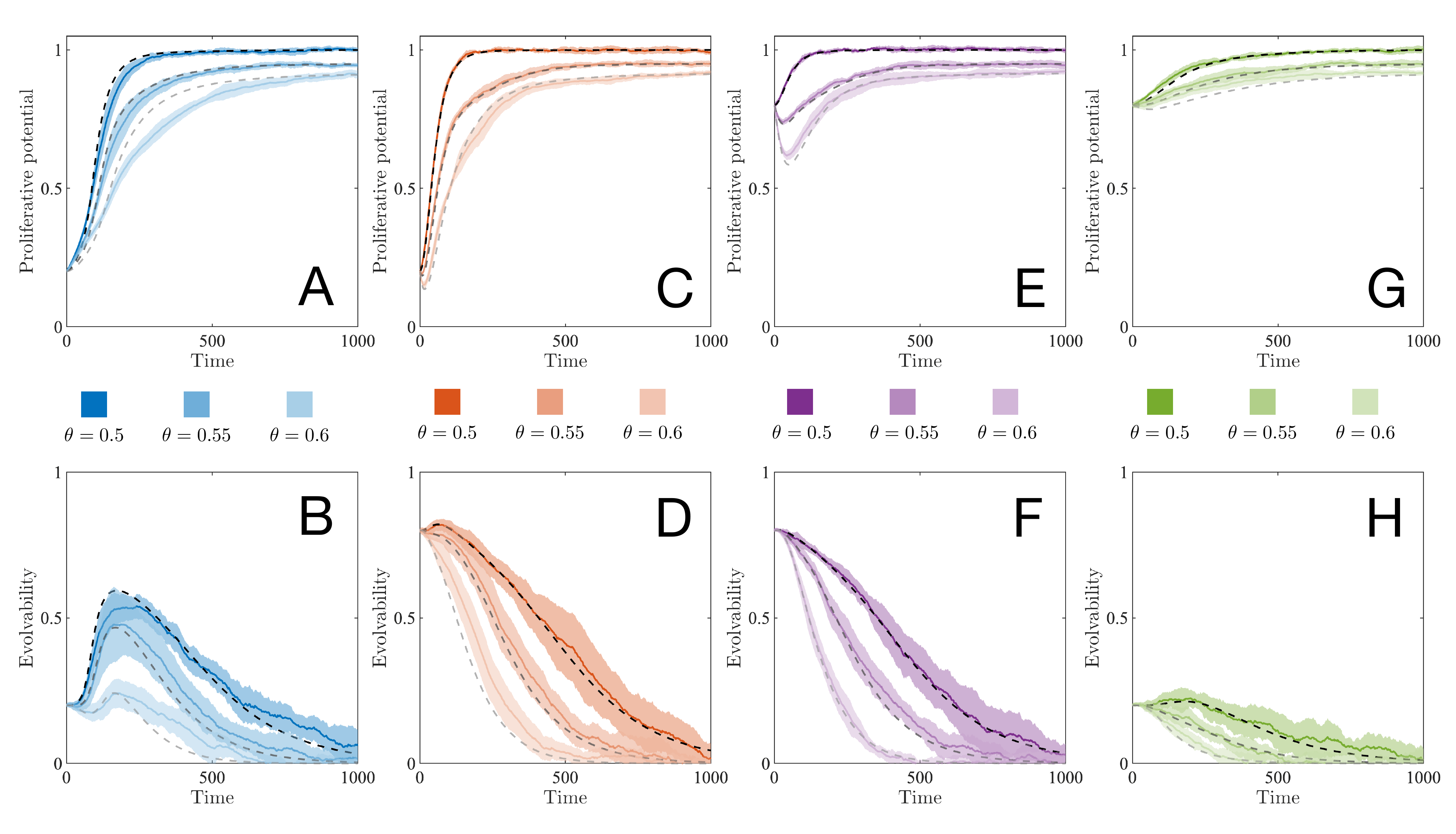}
\caption{{\bf How asymmetrical DFE affects the evolutionary dynamics.} Dynamics of the mean proliferative potential (panels \textbf{A}, \textbf{C}, \textbf{E}, and \textbf{G}) and the mean level of evolvability (panels \textbf{B}, \textbf{D}, \textbf{F}, and \textbf{H}). Solid coloured lines display the averaged results of numerical simulations of the IB model for different values of the initial mean levels of evolvability and proliferative potential -- i.e. $(\bar{x}^0, \bar{y}^0)=(0.2,0.2)$ (panels \textbf{A-B}), $(\bar{x}^0, \bar{y}^0)=(0.8,0.2)$ (panels \textbf{C-D}), $(\bar{x}^0, \bar{y}^0)=(0.8,0.8)$ (panels \textbf{E-F}), and $(\bar{x}^0, \bar{y}^0)=(0.2,0.8)$ (panels \textbf{G-H}). Dashed black lines display the results of numerical simulations of the continuum model. Lower transparency corresponds to a more asymmetrical DFE, with $\theta \in \{0.5, 0.55, 0.6\}$ (see legend). The results from the IB model correspond to the average over 10 simulations and the related standard deviation is displayed by the coloured areas surrounding the curves. Numerical simulations of the IB model were carried out using the initial phenotypic distribution defined via Eq.~\eqref{eq:ICIB} and the parameter values listed in Table~\ref{Table1} with $\alpha=0$ and $\eta=0.5$ in Eq.~\eqref{def:rho}. Details of the numerical simulations of the continuum model are provided in the Appendix.}
\label{Fig1revision}
\end{figure}

We first explore the influence of DFE asymmetry on the evolutionary dynamics of cell populations with different initial phenotypic distributions when $\alpha=0$. In Figure~\ref{Fig1revision} we can observe the mean levels of proliferative potential and evolvability of populations starting with low mean levels of both traits (cf. Figures~\ref{Fig1revision}{\bf A} and \ref{Fig1revision}{\bf B}), high mean levels of evolvability and low mean levels of proliferative potential (cf. Figures~\ref{Fig1revision}{\bf C} and \ref{Fig1revision}{\bf D}), high mean levels of both traits (cf. Figures~\ref{Fig1revision}{\bf E} and~\ref{Fig1revision}{\bf F}), and high mean levels of proliferative potential and low mean levels of evolvability (cf. Figures~\ref{Fig1revision}{\bf G} and \ref{Fig1revision}{\bf H}). A stronger DFE asymmetry (namely, a larger value of $\theta>0.5$) decreases the asymptotic level of proliferative potential to which the population converges in the long term and slows down the corresponding rate of convergence. Regarding evolvability, the transient phase whereby high evolvability is selected for observed when $\theta=0.5$ becomes negligible as $\theta$ increases above $0.5$. This appears to be the reason why the proliferative potential converges more slowly to a suboptimal value as $\theta$ increases: since cells with high levels of evolvability are now penalised due to the asymmetry in the DFE, cells in the population cannot rely on them to quickly traverse the phenotypic landscape in their search for the optimal proliferative potential. Still, in the long term, low evolvability is selected for regardless of the initial phenotypic distribution or the value of $\theta$.

\subsubsection{Asymmetrical DFE can promote population extinction}
In Subsection~\ref{sec:prevsec33}, we observed that populations exposed to very harsh environments (namely, large values of the parameters $\alpha$ and $\eta$) may become extinct in the IB model, due to the emergence of regions of the phenotypic space where the intrinsic net division rate $\rho(x,y)$ attains negative values. Here we investigate whether asymmetrical DFE can promote population extinction even if the intrinsic net division rate remains positive throughout the phenotypic space -- i.e. when $\eta<\gamma$ and $\alpha=0$ in the definition of the intrinsic net division rate provided by Eq.~\eqref{def:rho}.

\begin{figure}[hbtp]
\centering
\includegraphics[width=\textwidth]{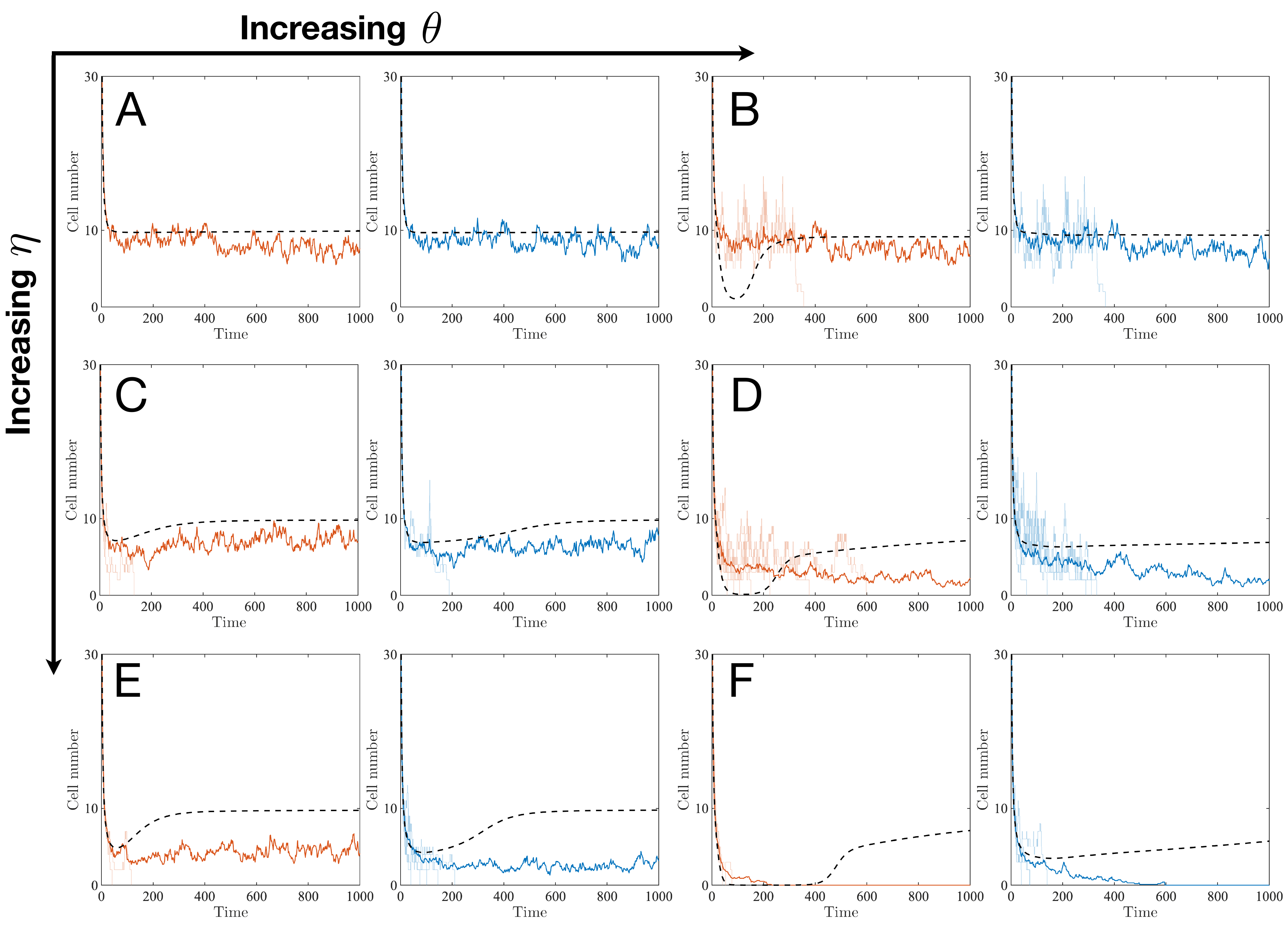}
\caption{{\bf How the interplay between asymmetrical DFE and the strength of natural selection affects the evolutionary dynamics.} Dynamics of the cell number. Panels with blue curves correspond to a low initial mean level of evolvability and a low initial mean level of proliferative potential -- i.e. $(\bar{x}^0, \bar{y}^0)=(0.2,0.2)$. Panels with orange curves correspond to a high initial mean level of evolvability and a low initial mean level of proliferative potential -- i.e. $(\bar{x}^0, \bar{y}^0)=(0.8,0.2)$. Solid coloured lines display the averaged results of 10 numerical simulations of the IB model. Solid transparent lines display the results of single simulations of the IB model where population extinction occurs (i.e. IB model simulations that ended with a cell number of 0). Dashed black lines display the results of numerical simulations of the continuum model. Numerical simulations of the IB model were carried out using the initial phenotypic distribution defined via Eq.~\eqref{eq:ICIB} and the parameter values listed in Table~\ref{Table1} with $\alpha=0$ in Eq.~\eqref{def:rho} but with $\gamma = 0.1$, $\kappa = 0.01$, and: $\theta = 0.5,\,\eta = 0.005$ (pair of panels {\bf A}), $\theta = 0.7,\,\eta = 0.005$ (pair of panels {\bf B}), $\theta = 0.5,\,\eta = 0.05$ (pair of panels {\bf C}), $\theta = 0.7,\,\eta = 0.05$ (pair of panels {\bf D}), $\theta = 0.5,\,\eta = 0.095$ (pair of panels {\bf E}), and $\theta = 0.7,\,\eta = 0.095$ (pair of panels {\bf F}). Details of numerical simulations of the continuum model are provided in the Appendix.}
\label{Fig2revision}
\end{figure}

We start by noticing that, when all population members have a positive intrinsic net proliferation rate, a drastic decline in size induced by overpopulation can be expected to be required for a population to become vulnerable to extinction due to demographic stochasticity~\cite{bouzat2010}. In the context of our model, under the definitions given by Eqs.~\eqref{def:R} and~\eqref{def:rho} with $\alpha=0$, the carrying capacity of the population is proportional to the parameter $\gamma$ and inversely proportional to the parameter $\kappa$. Hence, keeping the initial cell number fixed to the baseline value reported in Table~\ref{Table1}, we now choose values of $\gamma$ and $\kappa$ that are, respectively, smaller and larger than those in Table~\ref{Table1}, and we also reduce the value of $\eta$ accordingly so as to ensure that $\eta<\gamma$ and thus $\rho(x,y)>0$.  Under this set-up, we carry out numerical simulations of the IB and PIDE models for different combinations of $\theta$ and $\eta$, to study whether extinctions can occur in the IB model simulations. Results are displayed in Figure~\ref{Fig2revision}.

In Figures~\ref{Fig2revision}{\bf A} and \ref{Fig2revision}{\bf B}, since the selection gradient $\eta$ is relatively small, the fitness landscape has a low-slope, and the rate of adaptation is slower (similarly to Figure~\ref{Fig2}). When the DFE is symmetrical (cf. Figure~\ref{Fig2revision}{\bf A}), extinctions do not take place; on the contrary, when there is an asymmetry in the DFE (cf. Figure~\ref{Fig2revision}{\bf B}), extinctions may occur. In such case, the decline in cell number becomes more pronounced for populations with initially high evolvability (cf. the orange curves), hence providing the substrate for extinctions to be more likely in the IB model. In Figures~\ref{Fig2revision}{\bf C} and \ref{Fig2revision}{\bf D}, a relatively large selection gradient $\eta$ provides a steeper fitness landscape, triggering extinctions with a higher likelihood. Now, the population decline becomes more pronounced both in the symmetrical DFE (cf. Figure~\ref{Fig2revision}{\bf C}) and asymmetrical DFE (cf. Figure~\ref{Fig2revision}{\bf D}) cases. Moreover, in the latter case, populations starting with high evolvability (cf. the orange curves) are driven to a population bottleneck, as predicted by the PIDE model. Finally, in Figures~\ref{Fig2revision}{\bf E} and \ref{Fig2revision}{\bf F}, the steepness of the fitness landscape is even greater because an even larger selection gradient $\eta$ is considered. This leads to a faster adaptation, but also a faster (and sharper) decline in cell numbers. This is why, on average, populations become extinct sooner than before in the IB model simulations.

\subsubsection{How does the interplay between `endogenous' and `exogenous' costs of evolvability affect the evolutionary dynamics?}
Now we investigate scenarios where, in addition to the endogenous cost of evolvability associated with asymmetrical DFE, an exogenous cost associated with energy investment to keep low levels of evolvability can also be present~\cite{andre2006}. With this aim, keeping $\theta > 0.5$ (i.e. $\xi_3>0$), we define the intrinsic net division rate via Eqs.~\eqref{def:rho} and~\eqref{def:r2} with $\alpha > 0$, and carry out numerical simulations of the IB and PIDE models for different combinations of $\theta$ and $\alpha$.

\begin{figure}[hbtp]
\centering
\includegraphics[width=\textwidth]{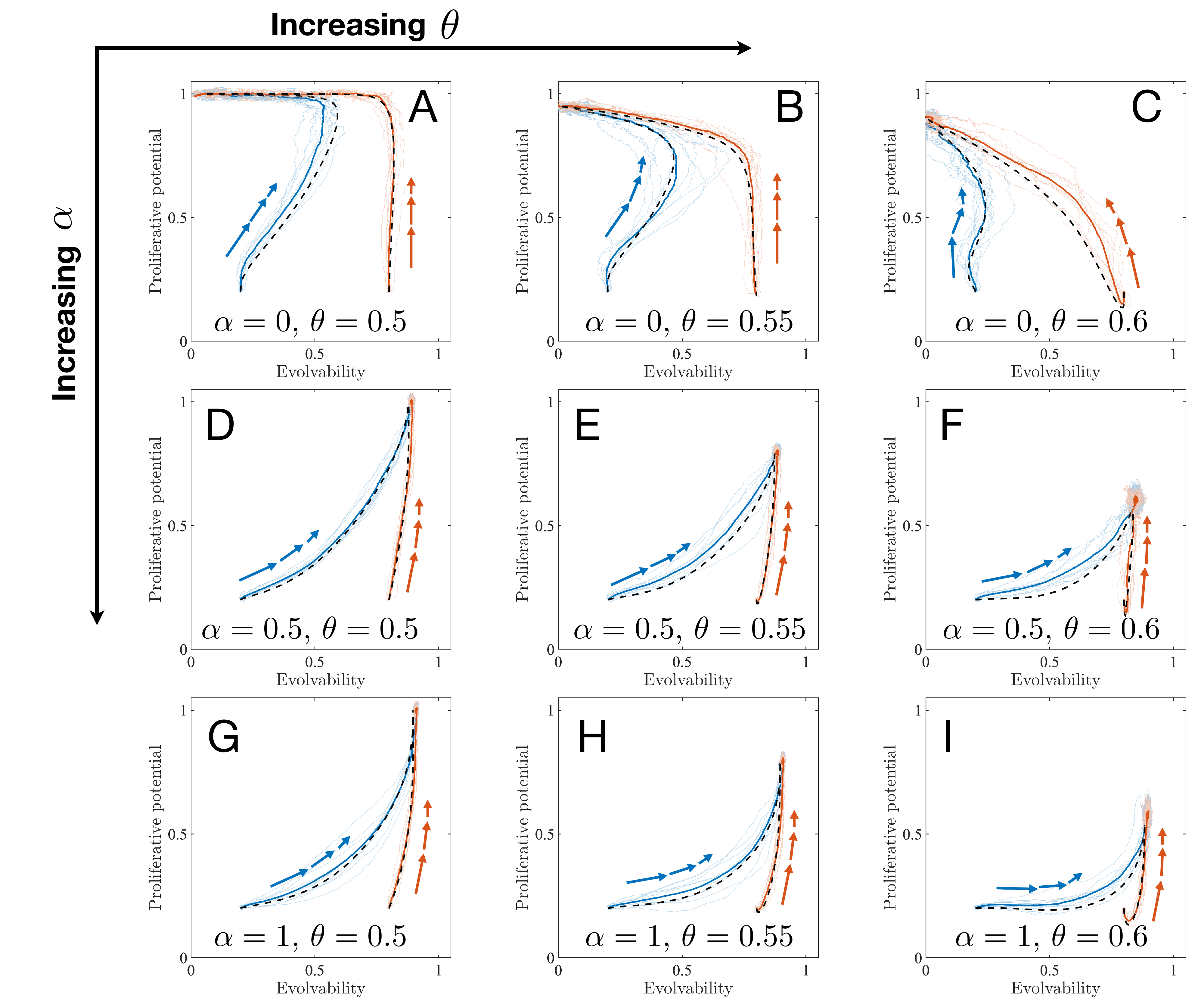}
\caption{{\bf How `endogenous' and `exogenous' costs of evolvability affect the evolutionary dynamics.} Dynamics of the mean phenotypic state for $t_{h} \in [0, 1000]$. Blue curves correspond to a low initial mean level of evolvability and a low initial mean level of proliferative potential -- i.e. $(\bar{x}^0, \bar{y}^0)=(0.2,0.2)$. Orange curves correspond to a high initial mean level of evolvability and a low initial mean level of proliferative potential -- i.e. $(\bar{x}^0, \bar{y}^0)=(0.8,0.2)$. Solid coloured lines display the averaged results of 10 numerical simulations of the IB model. Solid transparent lines display the results of single simulations of the IB model where population extinction occurs (i.e. IB model simulations that ended with a cell number of 0 ). Dashed black lines display the results of numerical simulations of the continuum model. The grey arrows in panel \textbf{C} indicate the direction of the phenotypic state trajectories. Numerical simulations of the IB model were carried out using the initial phenotypic distribution defined via Eq.~\eqref{eq:ICIB}, the definition of the intrinsic net division rate given by Eq.~\eqref{def:rho} complemented with Eq.~\eqref{def:r2}, and the parameter values listed in Table~\ref{Table1} with $\eta=0.5$ and the values of $\alpha$ and $\theta$ specified in the different panels. Details of numerical simulations of the continuum model are provided in the Appendix.}
\label{Fig3revision}
\end{figure}

The simulation results in Figure~\ref{Fig3revision} show how the evolutionary dynamics of the cell population change in this scenario. These results show that, in the absence of an exogenous cost over low levels of evolvability (i.e. when $\alpha = 0$), low evolvability cells are selected for, regardless of the value of $\theta$ (cf. Figures~\ref{Fig3revision}{\bf A}-\ref{Fig3revision}{\bf C}). On the contrary, under the effect of an exogenous cost over low levels of evolvability (i.e. when $\alpha > 0$), high evolvability cells are selected for (cf. Figures~\ref{Fig3revision}{\bf D}-\ref{Fig3revision}{\bf I}) in the long term, hence losing the evolutionary trend observed in the previous sections and in Figures~\ref{Fig3revision}{\bf A}-\ref{Fig3revision}{\bf C}. For increasing values of $\alpha > 0$ and $\theta > 0.5$, not only does the asymptotic proliferative potential becomes smaller (compared with the base case where $\alpha = 0$), but so does the asymptotic level of evolvability.

\section{Discussion and conclusions}
This theoretical study sheds light on the impact of evolvability on the evolutionary dynamics of phenotypically-structured cell populations. As a natural extension of other works in the field \cite{bukkuri2023,ardaseva2020,ardaseva2020b,ortega2022,macfarlane2022}, here we assume that heritable phenotypic variation and adaptive plasticity can be condensed into a cellular evolvability trait subject to evolution \cite{c4_evol2}. Thus our definition of evolvability lies between those of the \textit{heritability} and \textit{evolvability} (\textit{sensu} Wagner) concepts as defined in \cite{c4_evol3}. At a large evolutionary scale, evolvability could also be termed \textit{innovation}, generating major phenotypic (morphological, behavioural or physiological) breakthroughs \cite{c4_evol3, maynard1995}. That connotation of evolvability strays from the definition considered in this paper. 

The evolution of mutation rate and its determinants have been extensively explored from a mathematical point of view by other authors \cite{avila2023, xiong2008, goldi1979, andre2006, huang2017}. In this respect, it is relevant to point out that in this work \textit{evolvability} and \textit{mutation rates} are not synonyms. In fact, we do not focus solely on phenotypic changes driven by genetic mutations. Mutation rates are known to evolve, thus influencing the adaptive potential of organisms. Evolvability encompasses changes in mutation rates, but since it operates at several scales, it also encompasses changes in other mechanisms providing variability, such as phenotypic plasticity. 

Phenotypic plasticity can be an elusive concept, and precisely delineating its biological influences can be challenging. Nevertheless, it plays a pivotal role in the proper functioning of biological systems across various scales. At a cellular level, it is manifested through biological stochasticity, genetic and epigenetic diversity, and macromolecule stability regulation. On a broader scale, it is reflected in ecological phenomena such as niche partitioning, species interactions, and natural disturbances. The impact of phenotypic plasticity is intensified under harsh environments such as tumour development \cite{jenkinson2017} or perturbated ecosystems \cite{couzens2018, choi2019, pincheira2015}. Although it is difficult to relate theoretical predictions to real examples, it has been assumed that phenotypic plasticity is also subject to evolution \cite{masel2007,chevin2017,rago2019}.

It is also worth stressing that, despite the plethora of mechanisms contributing to it, evolvability is assumed to be a single quantitative cell trait in our model, modulating the ability of a cell to change its proliferative potential \cite{pienta2020}. Similar to other phenotypic traits, it is susceptible to spontaneous stochastic alterations in each cell, affecting its adaptive potential. Our way of implementing evolvability is subject to certain assumptions. The distribution of fitness effects (DFE) on cells undergoing changes in their proliferation potential (with a probability linked to their respective levels of evolvability) depends on a bias parameter $\theta$, which models the probability for the cells of acquiring a lower level of proliferative potential. We studied both the case where the DFE is symmetrical ($\theta = 0.5$) and cases where asymmetry in the DFE is present ($\theta > 0.5$). In either case, in the absence of any cost, evolvability is selected against after long periods of environmental stasis. As $\theta \rightarrow 0.5^+$, a transient phase where evolvability is selected for becomes relevant, helping cells traverse their phenotypic landscape more quickly. However, if we consider an exogenous cost for cells with low levels of evolvability, the opposite occurs: high evolvability is selected for in the long term. Although these results do not consider changing environments, it is known that, under such circumstances, evolvability can be linked to a suboptimal proliferative potential state \cite{masel2007, rago2019, perez2023}. By allowing different transition probabilities to nearby proliferative potential states, our model reproduced this phenomenon (Figure~\ref{Fig1revision}). However, future work should be conducted where changing environments are implemented in the context of this model, evaluating its influence on the adaptive dynamics of clonal populations. 
 
The interplay between proliferation rate and evolvability potential remains experimentally unknown. We approached this relationship from a qualitative point of view based on already known biological insights. The assumption that evolvability modulates the proliferative potential of a cell is based on the assertion that greater adaptability correlates with increased species success and is purposed as a first step in the study of increasingly complex mathematical models of evolutionary dynamics that include adaptive plasticity. As a starting point, we studied whether the outcome of the system evolution could be dependent on the initial phenotypic composition of the population, provided there is no fitness cost associated to evolvability, and that the DFE is symmetrical (as a base-case). We observed an asymptotic convergence of the simulations to the \textit{maximum proliferative potential -- lowest evolvability} phenotype regardless of the initial condition in an undisturbed environment (Figure~\ref{Fig2}). This is consistent with earlier studies indicating that low values of evolvability are positively selected in undisturbed environments and ensure balance and homeostasis at all levels of organisation \cite{chatterjee2017, lee2019, moore2020, cagan2022, lynch2011, zeldovich2007}.

However, in our analysis of the results at intermediate time points under various initial conditions, we observed that cells with a higher degree of evolvability appear to gain a competitive advantage during the initial stages of evolutionary dynamics. This advantageous trait enables them to navigate the phenotypic landscape more swiftly, thereby expediting the process of adaptation. These adaptive dynamics occur through a two-step process. First, cells with elevated evolvability are favoured in the short term. However, cells exhibiting high proliferative potential and lower evolvability are favoured in the long term, leading to a phenomenon of canalisation towards a robust phenotype \cite{waddington1957, wang2008}. When there is no fitness cost associated with evolvability and the initial mean level of evolvability of the cell population is sufficiently low, phenotypic variants with a relatively high evolvability level may have a temporary competitive advantage over the others on intermediate time scales. However, as soon as the mean level of proliferative potential of the cell population becomes sufficiently high, variants with high evolvability are outcompeted by lower evolvability variants.

The fundamental principles underlying seemingly different phenomena -- such as ecosystem regulation, which occurs on large physical and temporal scales, and tumour development, which unfolds over an organism's lifetime -- appear more similar than one might expect. At a broader scale, adaptive radiation \cite{gavrilets2009} or biodiversity hotspots \cite{madrinan2013, Orme2005, myers2000} represent highly evolvable scenarios \cite{gavrilets2005}. Adaptive radiation is usually followed by a decline phase in diversity which is linked to the continuous adaptation of resident niche specialists \cite{Meyer2011}. These dynamics relate the results of our population-level models with the \textit{less-evolvable more-proliferative} phenotype succeeding in the long term, and that evolvability levels are not constant in time. 

Following on this analysis in the case where there is no cost associated with evolvability, and the DFE is symmetrical, the influence of selection strength on proliferative potential was also studied, showing that the greater its intensity, the greater the competitive advantage of cells with high evolvability (during the early stages of selection). The results in Figure~\ref{Fig3} indicate that, when evolvability does not imply a fitness cost, a stronger selection on the proliferative potential speeds up the selective sweep underlying the fixation of fast-dividing phenotypic variants and it catalyses the selection of phenotypic variants with low evolvability on the long time scale. Thus, stronger selection strength favours or accelerates adaptation, as each jump in the direction of the optimal proliferative potential receives a greater reward. Under challenging environments, increased adaptive plasticity seems to be selected in the short-term in our simulation results, resembling what happens in other natural contexts including adaptive radiation and the onset of therapy resistance \cite{russo2024}. 
These dynamics mirror the negative epistasis phenomenon observed in numerous fitness landscapes. In this scenario, as a population becomes better adapted to its environment, the success of an advantageous mutation within that population becomes increasingly challenging. This has been reported in asexual \textit{E. coli} populations \cite{wiser2013} although average fitness (i.e. growth rate) continues to increase in time. Other authors reported that such a decrease in adaptability is best explained by the reduction of beneficial changes available in the phenotypic space in the same species \cite{wunsche2017}. 

This leads to another fact worth noting: high evolvability may not be \textit{free of charge} for the cell. In the aforementioned discussion about the case of symmetrical DFE we have ignored the deleterious effects of being evolvable and thus compromising the integrity of fundamental cellular mechanisms. However, it is known that beneficial mutations are less likely than detrimental ones \cite{eyre2007}; this implies that a greater level of evolvability may bring an endogenous cost that penalises cell viability. A high level of phenotypic plasticity might also compromise cell viability and ultimately phenotypic survival \cite{bukkuri2023,giraud2001,snell2010}. Likewise, a low mutation rate may require such an investment in keeping genome fidelity that proliferative potential may be adversely affected \cite{andre2006}. Hence, it is feasible to assume that evolvability implies a fitness cost. 

As a first approach to this issue, we considered a fitness cost for high levels of evolvability (Figure~\ref{Fig4}), which is given (in the context of the model) by values of $\alpha > 0$ in the intrinsic net division rate (as defined in Eq.~\eqref{def:rho} complemented with Eq.~\eqref{def:r1}). These numerical results support the conclusion that a greater fitness cost of evolvability may cause faster selection of phenotypic variants with low evolvability, thus slowing down the selective sweep that underlies the fixation of fast-dividing phenotypic variants. The inclusion of a cost of evolvability in the model hinders the selection of cells with higher evolvability during the early stages of evolutionary dynamics, but even in the absence of such an exogenous cost (i.e. when $\alpha=0$), an asymmetrical DFE also triggers such an effect (Figure~\ref{Fig1revision}). The asymmetry in the DFE produces more detrimental than beneficial variants; this encompasses an endogenous cost on cells with high levels of evolvability. Although the limits of plasticity have not been studied in as much depth as the benefits of maintaining phenotypic diversity, they have been shown to have evolutionary consequences \cite{dewitt1998}. Subsequently, cells with high evolvability may suffer from developmental instability and decreased robustness \cite{klingenberg2019, williams1957,  bien2019, scheiner1993}. In the context of the model, we are able to reproduce this by several means, as demonstrated in the discussion above.

Conversely, in Figure~\ref{Fig3revision}, we explore the scenario whereby low levels of evolvability are considered to incur a cost. Again, cost is increased by considering values of $\alpha > 0$ in the intrinsic net division rate (but this time, as defined by Eq.~\eqref{def:rho} complemented with Eq.~\eqref{def:r2}). There, the long-term dynamics change abruptly: now, cells with high levels of evolvability (but not the maximum attainable level of evolvability, since we consider an asymmetrical DFE) are selected for. If the cost effect on the proliferative potential is small (i.e. $\alpha \rightarrow 0^+$), the deviation from the base case is limited: evolvability is selected against in the long term, recapitulating previous theoretical and empirical results \cite{andre2006,sniegowski2000}. On the contrary, as the cost increases, evolvability is selected for, regardless of the initial phenotypic distribution of the population. This finding recapitulates the behaviour observed in some viruses, whereby high mutation rates are favoured for them to keep their adaptive capacity and high division rates \cite{furio2005,furio2007}.

In the context of cancer, tumour cells appear to have compensated for possible negative effects of higher evolvability through the parallel selection of alternative multiple metabolic and genetic mechanisms that enhance cell viability, such as cell redundancy, copy number variation, and degeneration. 
Therefore, tumour cells may be better able to inhabit this low-intermediate evolvability fitness cost window, leaving room for exploratory behaviour and leading to the known increased phenotypic heterogeneity observed in cancer \cite{Househam2021, Neftel2019, gupta2019, zhang2022}. 

Finally, we explored the occurrence of extinctions in the context of our model. Extinctions can take place by different means; in the analysis with symmetrical DFE, they may arise due to the impact of different phenotypic landscapes on the evolutionary dynamics of cell populations when both the strength of selection and the cost of evolvability come into play. In the most restrictive cases, where high selection strength and high cost of evolvability coincide, the phenotypic landscape can be so harsh as to create the conditions for the extinction of certain populations (depending on their initial phenotypic distribution) due to demographic stochasticity \cite{ardaseva2020, ardaseva2020b}. This phenomenon is similar to what occurs in threatened populations, where small groups (typically fewer than 1,000 individuals) face a high risk of extinction due to mutational meltdown -- an accumulation of harmful phenotypic traits -- over roughly 100 generations \cite{lynch1995, olofsson2023, zhang2022}, and it has even been imposed as a therapeutic strategy \cite{jensen2020}, especially in the context of antiviral therapy. However, we showed that extinctions may not only occur under such circumstances; they may also take place even when all cells in the population have a positive intrinsic net division rate. In Figure~\ref{Fig2revision}, we observe that, in scenarios where overpopulation induce a drastic decline in cell numbers in the early stages of population dynamics and selection gradients are sufficiently strong, the population may be stochastically driven to extinction too in the presence of an asymmetrical DFE,  especially when the initial evolvability of the population is sufficiently high.

In our modelling framework, such extinctions can only be observed using the discrete IB model, given its stochastic nature; in the continuum PIDE model, the population never goes extinct unless a population reaching a size smaller than 1 is considered to be extinct, such as in \cite{bukkuri2023}. However, when population bottlenecks occur in the PIDE model, extinctions in the IB model become more likely. This disagreement between the two models shows the importance of both paradigms when studying the evolutionary dynamics of cell populations: deterministic continuum models allow us to study the average behaviour of large populations, while stochastic discrete models allow us to observe rare phenomena that can be magnified at small population sizes. In future work, it would be insightful to perform a mathematical analysis of the continuum model used in this work, to explore the qualitative and quantitative properties of the solutions of this model, and to study the asymptotic behaviour of the cell populations described by it, evaluating the influence of evolvability in the evolutionary dynamics. 

The findings presented in Figures~\ref{Fig5},~\ref{Fig6}, and~\ref{Fig2revision} suggest that in more challenging environments, where stronger natural selection acts on proliferative potential, a substantial fitness cost associated with evolvability could potentially contribute to the extinction of cell populations. These cell populations would be primarily composed of slow-dividing, less fit phenotypic variants possessing a high degree of evolvability. The error catastrophe \cite{eigen1971} in viruses caused by mutation rates greater than a critical value is an example of such a situation, where a rapidly mutating viral genome loses the ability to preserve its integrity \cite{pariente2005}. The interaction between elevated values of $\alpha$ and $\eta$ (cf. Eq. \eqref{def:rho}) in our models appears to drive the population towards extinction due to demographic stochasticity. However, it is crucial to acknowledge that the persistence of a population is also contingent on genetic and phenotypic variance, and the facilitation of bounded adaptive evolution (evolvability) can promote the successful establishment of a population \cite{pienta2020, kanarek2010}.

In our current simple mechanistic model, we have focused on a static phenotypic landscape. Still, we have been able to recover well known evolutionary behaviours, as listed above. However, exploring the impact of evolving extrinsic selection pressures -- such as antibiotic, cytotoxic, or chemotherapeutic treatments, contingent on cell type -- on the evolutionary dynamics of populations undergoing changes in evolvability is a critical aspect for further investigation.
These considerations, including therapies affecting cell stemness or inducing chromosomal instability, may significantly influence the evolvability of cancer cells \cite{mcgranahan2012,yan2023}.
Extending our model to encompass various extrinsic pressures, such as conventional cytotoxic therapies affecting highly proliferative cells in cancer, but also potential epigenetic drugs targeting highly evolvable cells, could provide valuable insights that could support the development of innovative therapeutic approaches against cancer.

\backmatter

\bmhead{Funding}
This work was supported by the Spanish Ministerio de Ciencia e Innovación, the European Union NextGenerationEU/PRTR, MCIN/AEI/10.13039/501100011033 (grant numbers PID2022-142341OB-I00), Junta de Comunidades de Castilla-La Mancha (grant SBPLY/21/180501/ 000145) and by University of Castilla-La Mancha/European Regional Development Fund (FEDER; grant number 2022-GRIN-34405). JJS acknowledges funding support by Universidad de Castilla-La Mancha (grant number 2020-PREDUCLM-15634) and would like to thank also the Mathematical Institute (University of Oxford), and the Dipartimento di Scienze Matematiche (DISMA, Politecnico di Torino), for their support and hospitality during the development of this work. COS thanks the Spanish League Against Cancer (AECC) for their support (grant number 2019-PRED-28372). PKM would like to thank the Isaac Newton Institute for Mathematical Sciences, Cambridge, for support and hospitality during the programme Mathematics of Movement where work on this paper was undertaken. This work was supported by EPSRC grant no EP/R014604/1. 
TL gratefully acknowledges support from the Italian Ministry of University and Research (MUR) through the grant PRIN 2020 project (No. 2020JLWP23) “Integrated Mathematical Approaches to Socio-Epidemiological Dynamics” (CUP: E15F21005420006) and the grant PRIN2022-PNRR project (No. P2022Z7ZAJ) “A Unitary Mathematical Framework for Modelling Muscular Dystrophies” (CUP: E53D23018070001) funded by the European Union--NextGenerationEU. TL gratefully acknowledges also support from the Istituto Nazionale di Alta Matematica (INdAM) and the Gruppo Nazionale per la Fisica Matematica (GNFM).

\bmhead{Acknowledgements}
The authors would like to thank the two anonymous reviewers for their useful and insightful comments on the first version of the manuscript.

\bmhead{Competing interests}
The authors declare no competing interests.

\bmhead{Data availability}
This work does not rely on any data.

%\section*{Declarations}
%
%Some journals require declarations to be submitted in a standardised format. Please check the Instructions for Authors of the journal to which you are submitting to see if you need to complete this section. If yes, your manuscript must contain the following sections under the heading `Declarations':
%
%\begin{itemize}
%\item Conflict of interest/Competing interests (check journal-specific guidelines for which heading to use)
%\item Ethics approval and consent to participate
%\item Consent for publication
%\item Data availability 
%\end{itemize}
%
%\noindent
%If any of the sections are not relevant to your manuscript, please include the heading and write `Not applicable' for that section. 


\nolinenumbers

\renewcommand*{\bibfont}{\footnotesize}
%\bibliographystyle{plain}


\begin{appendices}

\small

\renewcommand{\thefigure}{A\arabic{figure}}
\setcounter{figure}{0}
\renewcommand{\theequation}{A\arabic{equation}}
\setcounter{equation}{0}
\renewcommand{\thesection}{A\arabic{section}}
\setcounter{section}{0}

\section*{Appendix}

\section{Formal derivation of the continuum model}
\label{sec:ApA1}

In the case where, between time-steps $h$ and $h+1$, each cell in phenotypic state $(y_i, x_j) \in (0,1) \times (0,1)$ undergoes phenotypic changes and divides or dies according to the rules underlying the IB model, the principle of mass balance gives the following difference equation

\begin{multline} \label{master}
n^{h+1}_{i,j} = \left(1 + \Delta t R(y_i, x_j, N^h) \right)  \\ \times \left\lbrace \mu(x_j) \left[ (1-\theta) n^h_{i-1,j} + \theta n^h_{i+1,j} \right] + \frac{\omega}{2} \left(n^h_{i,j-1} + n^h_{i,j+1} \right) \right.  + \left[1 - \left( \omega + \mu(x_j) \right) \right] n^h_{i,j} \biggr\}  .
\end{multline}
Using the fact that for $\Delta t$, $\Delta y$, and $\Delta x$ sufficiently small, the following relations hold
\begin{align*}
&t_h \approx t ,\hspace{5mm} t_{h+1} \approx t + \Delta t \\
&y_i \approx y ,\hspace{5mm} y_{i \pm 1} \approx y \pm \Delta y \\
&x_j \approx x ,\hspace{5mm} x_{j \pm 1} \approx x \pm \Delta x \\
&n^h_{i,j} \approx n(t,y,x) ,\hspace{4mm} n^{h+1}_{i,j} \approx n(t+\Delta t,y,x) \\
&n^{h}_{i\pm 1,j } \approx n(t, y \pm \Delta y, x) \\
&n^{h}_{i ,j\pm 1} \approx n(t, y, x \pm \Delta x) \\
&N^h \approx N(t) := \int_{0}^1\int_{0}^1 n(t,y,x)\; \textrm{d}y \;\textrm{d}x ,
\end{align*}
Eq.~\eqref{master} can be formally rewritten in the approximate form
\begin{multline} \label{cont}
n(t+\Delta t,y,x) = \left(1 + \Delta t R(y,x,N) \right) \times \left\lbrace \mu(x) \left[ (1-\theta) n(t,y-\Delta y, x) + \theta n(t,y+\Delta y,x) \right] \right. \\ \left. + \frac{\omega}{2} \left(n(t,y,x-\Delta x) + n(t,y,x+\Delta x) \right) \right.  + \left[1 - \left(\omega + \mu(x) \right) \right] n(t,y,x) \biggr\} .
\end{multline}
If $n(t,y,x)$ is a sufficiently regular function of $y$ and $x$ then for $\Delta y$ and $\Delta x$ sufficiently small we can use the Taylor expansions
\begin{equation*}
\begin{split}
n(t, y\pm \Delta y, x) = n(t,y,x) \pm \Delta y \partial_y n(t,y,x) + \frac{\left(\Delta y\right)^2}{2} \partial_{yy}^2 n(t,y,x)  + o(\left(\Delta y\right)^2)
\end{split}
\end{equation*}
and
\begin{equation*}
\begin{split}
n(t, y, x \pm \Delta x) = n(t,y,x) \pm \Delta x \partial_x n(t,y,x)  +  \frac{\left(\Delta x\right)^2}{2} \partial_{xx}^2 n(t,y,x) + o(\left(\Delta x\right)^2) \, .
\end{split}
\end{equation*}
Substituting these Taylor expansions into Eq.~\eqref{cont}, after a little algebra, we obtain
\begin{multline}
\dfrac{n(t+\Delta t, y, x) - n(t,y,x)}{\Delta t} = R(y, x, N) \, n(t,y,x) \\ + \left(\frac{\left(\Delta x\right)^2 \omega}{2 \Delta t} \partial_{xx}^2 n(t,y,x) + \frac{\left(\Delta y\right)^2 \mu(x)}{2 \Delta t} \partial_{yy}^2 n(t,y,x) + \frac{\Delta y (2\theta - 1) \mu(x)}{\Delta t} \partial_y n(t,y,x) \right) + {\rm h.o.t.} \, ,
\end{multline}
where higher order terms in $\Delta t$, $\Delta y$, and $\Delta x$ have been grouped into h.o.t. .
\\
If $n(t,y,x)$ is also a sufficiently regular function of $t$, letting $\Delta t \rightarrow 0^+$, $\Delta y\rightarrow 0^+$, $\Delta x\rightarrow0^+$, and $\theta \to 0.5^+$ in such a way that the conditions given by Eq.~\eqref{derived_parameters} are met, from the latter equation, rearranging terms, we formally obtain the PIDE~\eqref{eq:PIDE}. Finally, zero-Neumann (i.e. no-flux) boundary conditions on the boundary of the square $[0,1] \times [0,1]$ formally follow from the fact that the attempted phenotypic changes of the cells are aborted if they require moving into a phenotypic state that does not belong to the square $[0,1] \times [0,1]$.

\section{Details of numerical simulations of the continuum model}
\label{sec:ApA2}
To solve numerically the PIDE~\eqref{eq:PIDE} subject to no-flux boundary conditions on the square $[0,1] \times [0,1]$ and complemented with the continuum analogue of the initial condition defined via Eq.~\eqref{eq:ICIB}, i.e.

\begin{equation} \label{eq:ICPDE}
n(0,y,x) = n^0(y,x) := N^0 \, C \, \exp\left[-\frac{(y-\bar{y}^0)^2}{2(\sigma^0_y)^2}  - \frac{(x-\bar{x}^0)^2}{2(\sigma^0_x)^2}\right]
\end{equation}
where $C$ is a normalisation constant such that $\displaystyle{\int_0^1 \int_0^1 n^0(y,x) \, {\rm d}y \, {\rm d}x = N^0}$, we use a uniform discretisation of the interval $(0,1)$ as the computational domain of the independent variables $y$ and $x$, and a uniform discretisation of the interval $(0,t_f]$ with $t_f \in \left\{5\times 10^2,10^3\right\}$ as the computational domain of the independent variable $t$. The method for constructing numerical solutions is based on a three-point finite difference explicit scheme for the diffusion terms and an explicit finite difference scheme for the reaction term~\cite{leveque2007finite}. The parameter values are chosen to be consistent with those used to carry out numerical simulations of the IB model, which are specified in the main body of the paper. In particular, we define the values of $\xi_1$, $\xi_2$, and $\xi_3$ via Eq.~\eqref{derived_parameters}.

\end{appendices}

\end{document}